\documentclass[twocolumn]{aastex631}
\usepackage{CJK}
\usepackage{amsbsy}
\usepackage{multirow}
\usepackage{bm}
\usepackage{ulem}

\def\mpc{\,{\rm Mpc}}

\def\e{\,{\boldsymbol{e_1}}}

\def\dzab{\,{\rm \Delta Z_{AB}}}
\def\zrms{\,{\rm Z_{rms}}}
\def\dzabr{\,{\rm \Delta Z_{AB}^{random}}}
\def\dt{\,{\rm Z_{rms}/\Delta Z_{AB}}}

\received{XXX}
\revised{YYY}
\accepted{ZZZ}
\published{HHH}

\submitjournal{ApJ}

\shorttitle{filament spin}
\shortauthors{Tang \& Wang}

\begin{document}
\begin{CJK*}{UTF8}{gbsn}
\title{Cosmic filament spin - I: a comparative study in observation}

\correspondingauthor{Peng Wang}
\email{pwang@shao.ac.cn}

\author[0009-0001-7527-4116]{Xiao-xiao Tang(唐潇潇)}
\affil{Shanghai Astronomical Observatory, Chinese Academy of Sciences, Nandan Road 80, Shanghai 200030, People's Republic of China.}
\affil{University of Chinese Academy of Sciences, Beijing 100049, People's Republic of China.}

\author[0000-0003-2504-3835]{Peng Wang(王鹏)}
\affil{Shanghai Astronomical Observatory, Chinese Academy of Sciences, Nandan Road 80, Shanghai 200030, People's Republic of China.}

\author{Wei Wang(王伟)}
\affil{Purple Mountain Observatory, Chinese Academy of Sciences, No.10 Yuan Hua Road, 210034 Nanjing, People's Republic of China.}
\affil{School of Astronomy and Space Science, University of Science and Technology of China, Hefei 230026, Anhui, People's Republic of China.}
\affil{Shanghai Astronomical Observatory, Chinese Academy of Sciences, Nandan Road 80, Shanghai 200030, People's Republic of China.}

\author[0000-0002-9891-338X]{Ming-Jie Sheng(盛明捷)}
\affil{Department of Astronomy, Xiamen University, Xiamen, Fujian 361005, People's Republic of China.}

\author[0000-0001-5277-4882]{Hao-Ran Yu(于浩然)}
\affil{Department of Astronomy, Xiamen University, Xiamen, Fujian 361005, People's Republic of China.}

\author[0000-0003-1132-8258]{Haojie Xu(许浩杰)}
\affil{Shanghai Astronomical Observatory, Chinese Academy of Sciences, Nandan Road 80, Shanghai 200030, People's Republic of China.}


\begin{abstract}
In the cosmic web, filaments play a crucial role in connecting walls to clusters and also act as an important stage for galaxy formation and evolution. Recent observational studies claim that filaments have spin. In this study, we examined the potential impact of diversity in filament identification algorithms and galaxy survey datasets on the quantification of filament spin. The results of this study demonstrate qualitative agreement with previous research, suggesting that a reliable filament spin signal is detectable when the viewing angle of filament spine larger than 80 degrees under a rough estimation. The detected filament spin signal is intricately linked to the viewing angle, dynamic temperature, etc. The quantitative difference of filament spin signal among samples is slightly dependent on the filament identification algorithms, while the value is relatively greater dependent on the redshift space distortion effect in the galaxy sample.
\end{abstract}



\keywords{
    \href{http://astrothesaurus.org/uat/902}{Large-scale structure of the universe (902)};
    \href{http://astrothesaurus.org/uat/330}{Cosmic web (330)};
    \href{http://astrothesaurus.org/uat/2029}{Galaxy environments (2029)};
    \href{http://astrothesaurus.org/uat/1882}{Astrostatistics (1882)}}



\section{Introduction}
\label{sec:intro}

The modern standard $\Lambda$CDM cosmological model suggests that galaxies are not randomly distributed on the large scale of the Universe but instead form the cosmic web \citep{1996Natur.380..603B}, which is composed of voids, walls, filaments, and clusters. It has been confirmed by numerous galaxy surveys (such as the 2dF Galaxy Readshift Survey \citep{10.1046/j.1365-8711.2001.04902.x}, Sloan Digital Sky Survey \citep{2004ApJ...606..702T}) and simulations (such as EAGLE \citep{2015MNRAS.446..521S}, illustris \citep{2015A&C....13...12N}, IllustrisTNG \citep{2018MNRAS.475..676S}). Among the cosmic web, filaments play a crucial role in connecting walls to clusters and also act as an important stage for galaxy formation and evolution. 
Galaxies tend to form within the cosmic filaments, where the density of dark matter and gas is higher. Studies have shown that the properties of galaxies, such as their morphology \citep{2009ApJ...706..747Z}, star formation rates \citep{2012ApJ...749L..43C,Darvish_2014}, and shape and spin orientation \citep{2009ApJ...706..747Z,2015ApJ...798...17Z,Tempel_2013,2017MNRAS.468L.123W,2018MNRAS.473.1562W}, are influenced by their reside cosmic filaments.

A variety of methods have been developed to identify and classify cosmic structures such as filaments, clusters, and large-scale features in the cosmic web. These methods use diverse approaches, including graph-based algorithms (e.g., the Minimal Spanning Tree method, \citealt{2014MNRAS.438..177A}), stochastic modeling (e.g., Bisous, \citealt{2014MNRAS.438.3465T, 2016A&C....16...17T}; FINE, \citealt{2010MNRAS.407.1449G}), and tensor-based Hessian analysis (e.g., T-web, \citealt{2009MNRAS.396.1815F}; V-web, \citealt{hoffman2012kinematic}; CLASSIC, \citealt{kitaura2012linearization}). Advanced scale-space techniques are utilized by NEXUS+ (\citealt{2013MNRAS.429.1286C}) and MMF-2 (\citealt{2007ApJ...655L...5A, AragonYang2014}), while topological approaches are adopted by Spineweb (\citealt{2010MNRAS.408.2163A}) and DisPerSE (\citealt{2011MNRAS.414..350S}). Phase-space analyses are used in ORIGAMI (\citealt{falck2012origami, falck2015persistent}) and MSWA (\citealt{ramachandra2015multi}).
These methods vary in the types of structures they identify, such as filaments, clusters, or the entire cosmic web, and in their sensitivity to different datasets, including particles and galaxy/haloes. This variety highlights ongoing efforts to improve cosmic structure detection and advance our understanding of large-scale structure formation. However, the complexity of filamentary structures results in inconsistencies among the outcomes of different algorithms. Currently, there is no universally accepted method for defining filaments. For a detailed review of these classification techniques, we refer readers to \citep{2018MNRAS.473.1195L}.

The investigation revealed that the statistical properties of cosmic filaments are significantly affected by the choice of identification algorithm, as demonstrated in \citep{2020MNRAS.493.1936R}. This was shown by comparing three sets of filament catalogues generated by different algorithms applied to the Sloan Digital Sky Survey. The research found variations in statistical features, such as the redshift distribution and length, across different catalog sets. This suggests that the chosen algorithm can lead to differences in filament characteristics, potentially affecting the interpretation of cosmological effects. Consequently, this finding underscores the necessity of performing comprehensive analyzes on filament catalogs developed using various algorithms or data sets for a detailed study of filaments.

The investigation of the properties of the filaments themselves has gained significant attention. For example, studies such as \cite{2005MNRAS.359..272C, 2006MNRAS.370..656D, 2010ApJ...723..364A, 2010MNRAS.408.2163A, 2010MNRAS.407.1449G, 2010MNRAS.409..156B, 2014MNRAS.441.2923C, 2020A&A...638A..75B, 2020A&A...641A.173G, 2020A&A...637A..41T} have analyzed the density distributions of mass or galaxy numbers surrounding filament spines. \cite{2024MNRAS.532.4604W} assessed the galaxy density gradient to determine the radii of the filaments, finding approximately 1 Mpc at $z=0$ from both state-of-the-art hydrodynamical simulations and observational data. Recently, \cite{2021NatAs...5..839W} offered observational evidence of the rotation of cosmic filaments, demonstrating that cosmic angular momentum can be generated on unprecedented scales.

The filament spin, which is our main research interest, in a cosmological context remains a significant issue in cosmology. According to the standard model of structure formation, tiny initial over-density regions in the very early Universe expand through gravitational instability as matter moves from less dense to more dense areas. This mass flow is laminar, indicating the absence of primordial rotation in the early universe, so that angular momentum must form during the formation of structures. Tidal torque theory \citep{1969ApJ...155..393P, 1970Afz.....6..581D, 1970Ap......6..320D, 1984ApJ...286...38W,1987ApJ...319..575B, 1993ApJ...418..544V, 1996ApJS..103....1B, 1999A&A...343..663P, 2002MNRAS.332..325P} explains how misalignment between the inertia tensor of a collapsing space region \citep{1969ApJ...155..393P, 1970Ap......6..320D, 1999A&A...343..663P} and the tidal(shear) field can spin up the collapsing material. This explanation applies when density perturbations are small, known as the linear regime.

Following the discovery of possible evidence for filament spin in \cite{2021NatAs...5..839W}, several key tasks are essential to advancing our understanding of this phenomenon: 
\begin{enumerate}
    \item Reconfirming the signals of filament spin: It is essential to verify the filament spin signal, especially in light of the concerns raised in this manuscript. Specifically, we aim to determine whether consistent signals, at least qualitatively, can be obtained across different galaxy datasets and filament identification methods. This forms the primary motivation for our study.
    \item Replicating the observed spin signal in numerical simulations: Extending the analysis to simulations is a logical next step to better understand the differences between real and redshift space.
    \item Investigating the physical origin of filament spin: Finally, we need to delve deeper into the underlying physical formation and evolution of the filament spin.
\end{enumerate}
To address these tasks, we conducted a series of studies. As the first paper in our series on the topic of \textit{Cosmic Filament Spin}, we performed a comparative study of the filament spin signal. Specifically, we examined the influence of different filament identification algorithms and galaxy survey datasets, revisited the analysis of \cite{2021NatAs...5..839W} using identical methods, and investigated whether filament spin signals vary across algorithms, datasets, or filament catalogue samples.

The structure of the paper is as follows. Section~\ref{sec:method} outlines four sets of filament catalogues derived from various algorithms and their corresponding galaxy sample. Section~\ref{sec:result} discusses the spin signals discovered in these catalogues and compares the results. Finally, Section~\ref{sec:sum_dis} provides conclusions and a discussion of the results.

\section{Data and Methodology}
\label{sec:method}

This part outlines the galaxy and filament samples detailed in Section~\ref{sec:gal_sample} and Section~\ref{sec:fila_sample}, respectively.
It noted that all samples are derived from observed SDSS \citep{2015ApJS..219...12A} galaxy data.
The method used to quantify filament spin, as discussed in Section~\ref{subsec:Spin}.

\begin{table*}[!ht]
    \centering
    \caption{\textbf{Overview of the galaxy and filament catalogues used in this study.} Each column, from left to right, contains the referenced sample name, tracer galaxy, filament finder, number of filaments, galaxy employed to determine filament spin (measure galaxy), and the number of galaxies. In `Sample-2' and `Sample-3', the redshifts of galaxies in Measure Galaxy \citep{2015ApJS..219...12A} are not corrected for redshift space distortion (RSD).}
    \begin{tabular}{lccccccc}
         \hline
         \hline
          Ref.     & Tracer Galaxy & Filament Finder & Filament No. & Measure Galaxy & Gal. No.\\
         \hline
         Sample-1  &  \cite{2014MNRAS.438.3465T} & Bisous \citep{Stoica2005}  & 15,421  & \cite{Tempel2012} & 576,493 \\
         Sample-2  &  \cite{2015ApJS..219...12A} & MST \citep{2020MNRAS.499.4876P} & 8,350 & \cite{2015ApJS..219...12A}, RSD & \multirow{2}{*}{584,434} \\
         Sample-3  &  \cite{2011MNRAS.415.2553Z}  & M16 \citep{10.1093/mnras/stv2295} & 3,094 & \cite{2015ApJS..219...12A}, RSD &  \\
         Sample-4  &  \cite{2015ApJS..219...12A} & DisPerSE \citep{2011MNRAS.414..350S} & 26,974 & \cite{Tempel2017groups} & 584,449  \\
         \hline
    \end{tabular}
    \label{table:table1}
\end{table*}

\begin{figure*}[!ht]
\centering{\includegraphics[width=1.0\textwidth]{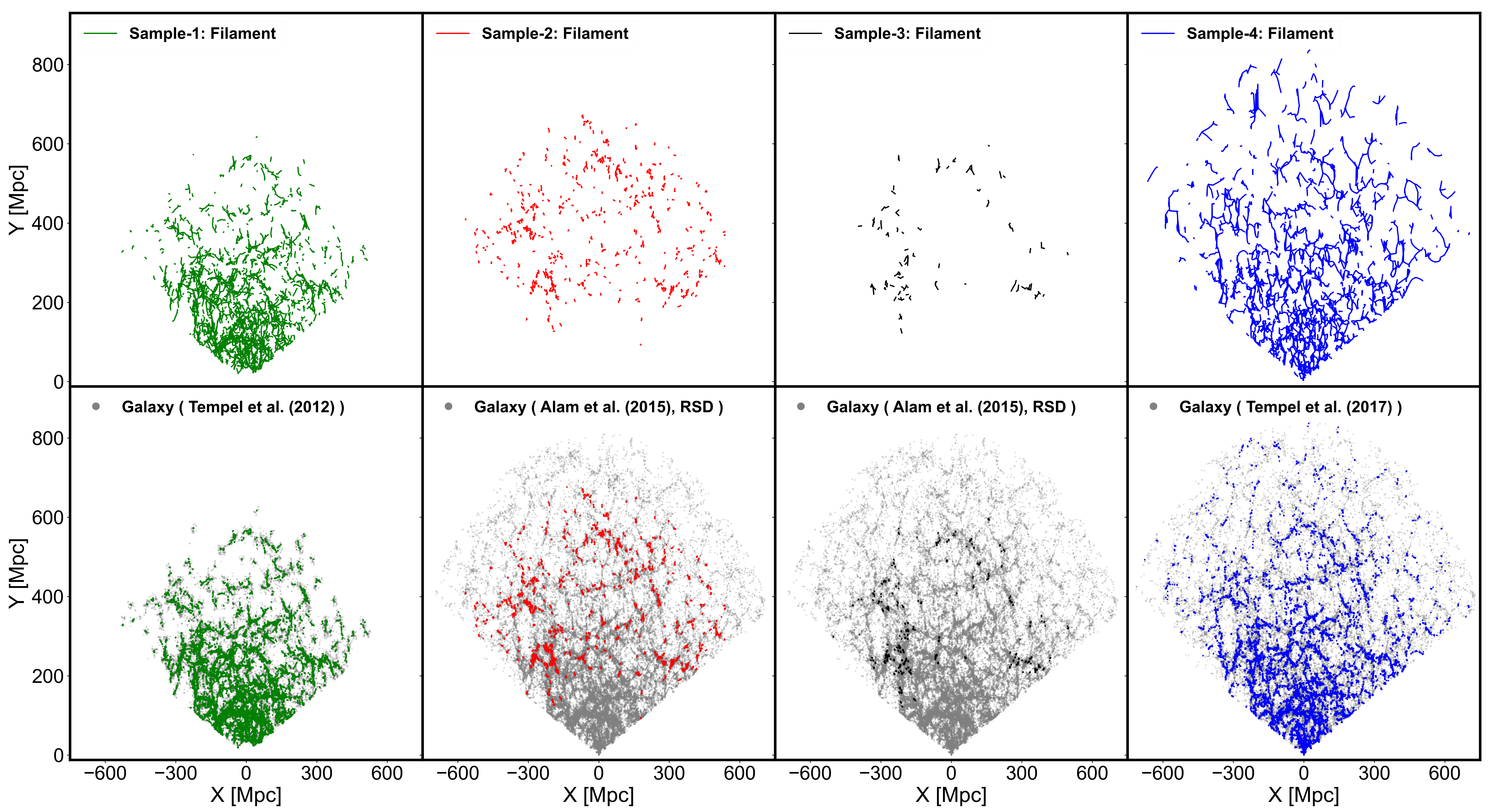}}
\caption{\textbf{The spatial distribution of filaments and galaxies.} Upper panels: The distribution of filaments before the selection. Bottom panel: the distribution of `measure galaxy' (grey dots in the background) and selected filaments. To compare the filaments produced by the four algorithms, the XY-axes have been normalized to the same range. Filaments and galaxies within range of $0<z<40$ $\mpc$ are displayed. }
\label{fig:fila_dis}
\end{figure*}

\begin{figure}[!ht]
\centering{\includegraphics[width=0.45\textwidth]{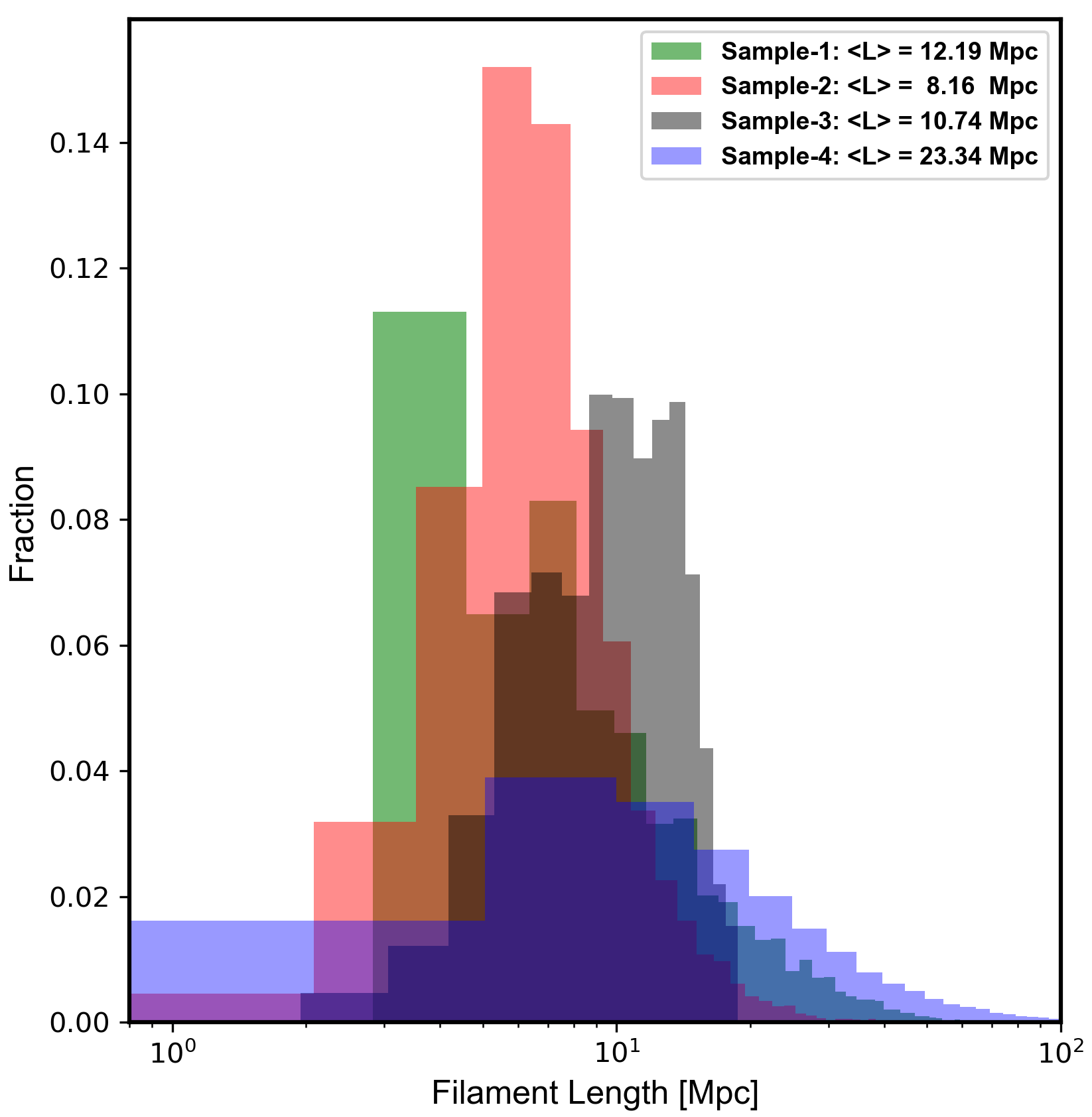}}
\caption{\textbf{The length distribution of filaments in four samples.} The mean value of the filament length for the corresponding sample is indicated in the legend.}
\label{fig:fila_len}
\end{figure}

\subsection{Filament samples}\label{sec:fila_sample}

As highlighted in \cite{2018MNRAS.473.1195L}, there are various algorithms available for the classification of cosmic filaments. To investigate whether and how the filament spin signal changes with different filament algorithms, we used four publicly available filament catalogs, as illustrated in Table~\ref{table:table1}, to compare the filament spin signal. It should be noted that the galaxy sample used to define filaments is referred to as `tracer galaxy' (the second column), while the galaxy sample used to measure the filament spin signal is named `measure galaxy' (the fifth column). 
Due to the methodology of the filament definition and the purpose of the investigation, the `tracer galaxy' in each sample are different.
From Table~\ref{table:table1}, we can find that the galaxy sample of `measure galaxy' is not exactly the same. It will be introduced in Section~\ref{sec:gal_sample}.  The galaxy and filament samples in `Sample-1', `Sample-2', and `Sample-3' are obtained from \cite{2020MNRAS.493.1936R}.

To visually show the differences of the filaments in these four filament catalogues, we chose the filaments located with a coordinate range of $0<z<40$, in units of $\mpc$, and their spatial distributions are shown in Figure~\ref{fig:fila_dis}. The length distribution of all filaments is illustrated in Figure~\ref{fig:fila_len}. The differences between the samples are quite evident: First, there is a distinct variance in their spatial distributions. Filaments in Sample-1 and Sample-4 show a relatively broader distribution, whereas Sample-2 and Sample-3 display a more clustered distributions. Second, there are also variations in the filament length distribution, as presented in both Figure~\ref{fig:fila_dis} and Figure~\ref{fig:fila_len}, where the filaments in Sample-2 are generally shorter than $10 \mpc$, while the filaments in Sample-3 are generally longer than $10 \mpc$, in contrast to the numerous longer filaments (see the mean value listed in the legend in Figure~\ref{fig:fila_len}) in Sample-1 and Sample-4. These differences primarily originate from the different tracer galaxies employed to define the filaments, as well as differences in the filament identified algorithms. In the following, a brief overview of the filaments within these four samples is given.

\subsubsection{Sample-1}
Filaments in the `Sample-1' (the first raw in the Table~\ref{table:table1}) are detected via the Bisous process \citep{Stoica2005}, which characterises the filament as an approximation consisting of several cylinders aligned along its axis, based on the spatial distribution of galaxies. A filamentary structure is constructed by analysing the connectivity and alignment of small, random segments (cylinders) derived from the spatial positions of galaxies. The spine of a filament can subsequently be identified on the basis of the probability of detection and the filament's orientation. We refer the reader to \cite{2014MNRAS.438.3465T, 2016A&C....16...17T} for further details and to \citep{2018MNRAS.473.1195L} for a comparison of the Bisous model with other popular cosmic web classification methods. The model has also been successfully applied to observational data and simulation catalogues \citep{2007JRSSC..56....1S,2010A&A...510A..38S}. 
The Bisous filament finder is the same as the observational work of the first measurement of filament spin \citep{2021NatAs...5..839W}.  Information about the filament lengths, the closest galaxy IDs and the distances is recorded. `Sample-1' comprises 15,421 filaments for subsequent analysis.

\subsubsection{Sample-2}
The second catalogue of filaments, referred to as `Sample-2', used in this study is derived from \citep{2020MNRAS.499.4876P}. It is constructed using a cosmological filament identification algorithm, which employs a technique from graph theory known as the minimal spanning tree \citep[MST,][]{1985MNRAS.216...17B}. The MST is an algorithm in graph theory that is used to find the spanning tree with the smallest total weight in a weighted graph. Based on previous investigations \citep{1985MNRAS.216...17B,2007MNRAS.375..337C,2014MNRAS.438..177A}, \cite{2020MNRAS.499.4876P} adopted the same definitions for identifying filaments.

In this context, a graph is made up of vertices (representing galaxies in observations), edges (which connect the vertices), and weights. The Minimum Spanning Tree (MST) is a distinct set of edges that connects all vertices in the graph without forming any closed loops and ensures the minimum total weight, provided all weights differ. 
The MST mainly illustrates the distribution of nearby neighbors, but tends to diffuse when dealing with many vertices, failing to accurately depict the large-scale structure. To overcome this, \citep{2020MNRAS.499.4876P} confine the MST to an intermediate density region, identified by a Friends-of-Friends (FoF) algorithm tailored for flux-limited galaxy surveys, as described by \citep{2005ApJ...630..759M}. 
\citep{2020MNRAS.499.4876P} utilize specific link lengths and edge weightings based on galaxy luminosities to form the MST using galaxies from SDSS DR12 \citep{2015ApJS..219...12A}. 
Branches of the tree ending with galaxies of brightness $M_r < -21.0$ were identified as filaments, resulting in a catalogue containing 8,350 filaments. The catalogue includes the physical length, elongation, galaxy count per filament, filament IDs, and the either spectroscopic or photometric redshifts of galaxies.

\subsubsection{Sample-3}
In `Sample-3', filaments are identified from a galaxy group catalogue \citep{2011MNRAS.415.2553Z}.
The filament catalogue \citep{10.1093/mnras/stv2295} was created considering pairs of groups, including only those with virial masses larger than $\rm M_{vir}>10^{13.5}$, within the redshift interval $0.05<z<0.15$. 
Filaments are represented as straight lines connecting nearby galaxy groups, thus inherently ignoring any curved or branched formations. Three selection criteria are applied: A pair of groups is connected by a filament if:

(a) Their radial velocity difference is less than $\rm 1000 \ km/s$; 

(b) Their projected separation is less than $\rm 10 \ Mpc/h$; 

(c) The galaxy overdensity in the region between the group pairs surpasses a certain threshold.

Clearly, based on the definition of filaments, this filament sample is incomplete. Nevertheless, \citep{10.1093/mnras/stv2295} aim to investigate the impact of filaments on galaxies that are falling into groups.  
Following the identification process, a catalogue was compiled, listing 3,094 filaments along with their physical characteristics, including the distance between groups (filament length), the count of member galaxies, and more, for subsequent studies.

\subsubsection{Sample-4}
Filaments in the `Sample-4' used in this study employ the discrete persistent structure extractor \citep[hereafter \texttt{DisPerSE;}][]{2011MNRAS.414..350S,2011MNRAS.414..384S} -- one of the most commonly employed cosmic web finders -- to detect cosmic filaments. The underlying density field is estimated using the Delaunay tessellation field estimator \citep[DTFE,][]{2000A&A...363L..29S, 2009LNP...665..291V} based on the galaxy distribution. The DTFE densities are smoothed to reduce the shot noise. 
Filaments are defined as segments that connect high-density critical points to saddle points along the density ridges. The significance level of filament persistence can be adjusted by modifying the density ratio of the two critical points in the pair \citep{2011MNRAS.414..350S,2011MNRAS.414..384S}. The impact of various \texttt{DisPerSE} parameters has been rigorously investigated in previous studies \citep[see e.g.,][]{2020A&A...634A..30M, 2020A&A...642A..19M,Galarraga2024}. We follow the parameters setting recommended by \cite{Galarraga2024} for filament extraction. In summary, we apply a single smoothing to the density field of DTFE and set the persistence significance level at 2$\sigma$. As shown in Table~\ref{table:table1}, `Sample-4' consists of 26,974 filaments for further analysis.

\subsection{The Measure Galaxy samples}\label{sec:gal_sample}
To address how the filament spin signal is influenced by galaxy samples, we used three galaxy samples for spin signal measurements, as listed in the last two columns of Table~\ref{table:table1}. The first galaxy sample, which was only used in the `Sample-1' comes from \cite{Tempel2012}, contains 576,493 galaxies. 
The galaxy sample used in `Sample-2' and `Sample-3' is the same, constructed from \cite{2015ApJS..219...12A}, contains 584,434 galaxies. 
The third galaxy sample \citep{Tempel2017groups}, used for `Sample-4' and identical to the galaxy sample in \cite{2021NatAs...5..839W}, was derived from a publicly accessible catalogue \citep{Tempel2017groups} and was compiled from \citep{2011AJ....142...72E, 2015ApJS..219...12A}, contains 584,449 galaxies.

Actually, all three galaxy samples (the galaxies overlap by approximately 90\%) were constructed from the Sloan Digital Sky Survey (SDSS). Galaxies are selected from the primary regions of the survey (Legacy Survey), and only galaxies with spectroscopic redshift are considered.  To ensure completeness of the spectroscopic sample and minimi se the influence of local superclusters, the lower magnitude limit of the sample is set at $M_r = 17.77$. The galaxy samples provide data on redshift, equatorial coordinates, etc., for filament detection and the measurement of the filament spin signal (refer to Section~\ref{subsec:Spin}). 

In observational data, the precise positions of galaxies, which are often derived from observed redshifts, are influenced by redshift space distortions (RSD). These distortions arise from the peculiar velocities of galaxies. In this work, we use the term `redshift-space distortions' (RSD) to refer to any effect that produces a difference between real and redshift space. This includes not only the Kaiser effect (linear-theory, large-scale RSD) but also smaller-scale effects such as the Fingers of God and the spin signal discussed here. For some of the samples analyzed, only the Kaiser effect was estimated and removed, leaving other RSD contributions intact. This partial removal ensures that the rotating filament signal remains detectable.

In our study, the redshifts of galaxies in `Sample-2' and `Sample-3’ are uncorrected for RSD, whereas those in `Sample-1' and `Sample-4’ are corrected. To mitigate the impact of RSD, \cite{Tempel2017groups} provides an effective approach. This approach involves associating galaxies from the group catalogue with groups identified using a modified Friends-of-Friends (FoF) halo finder \citep{1982ApJ...257..423H, 1985ApJ...292..371D, 2001MNRAS.328..726S, 2009MNRAS.399..497D}. Subsequently, corrections are applied to galaxy positions along the line of sight, taking into account both the radial velocity dispersion of the group and its angular size on the sky plane.

As mentioned before, the four filaments are identified with different `tracer galaxy' (see the second column in the Table~\ref{table:table1}), hence we must assign the galaxy sample of `measure galaxy' (the fifth column) to the filaments before analysing the filament spin. Each galaxy was linked to the closest filament according to their spatial locations. During this assignment, the galaxy distance to the filament spine and the filament ID were recorded. Following the suggestion of \cite{2021NatAs...5..839W}, only galaxies within a radius of 2 $\mpc$ from the filament spine were considered to contribute the spin signal. Consequently, we only consider galaxies within a distance of less than 2 $\mpc$.

The spatial distribution of our galaxy sample (the grey dots) in the sky is shown in the bottom of the Figure~\ref{fig:fila_dis}. For clarity, we only show galaxies within a slice of width $0<z<40$ $\mpc$ along the z-axis and within a distance less than 2 $\mpc$ to the spines of selected filaments. It is clear to see that the galaxy distribution shows some filamentary structure, which matches the filament distribution.

\subsection{filament spin}\label{subsec:Spin}
To study the spin of the filament in four samples and to compare our results with those found by \cite{2021NatAs...5..839W}, we used the same method to calculate the spin of the filament. The method is described in the following.

\cite{2021NatAs...5..839W} applied an new method to assess the spin of the filament in observations. Following the experience from the measurement of the rotation of spiral galaxies in observation by using the Doppler redshift or blueshift of the member stars, the filament's spin can accordingly be confirmed by examining the redshift or blueshift of the galaxies residing within it.
After assigning galaxies to their nearest filaments, we can calculate the average redshift $Z_{0}$ of all galaxies within a given filament. Galaxies in a given filament were then divided into two regions (A and B) on either side, and the mean redshifts of galaxies in either region A or B are calculated, denoted as $\rm Z_{A}$ and $\rm Z_{B}$, respectively.  We propose that if filaments are rotating, a significant component of a galaxy's velocity perpendicular to the filament's axis should be observable. This component should show opposite directions on either side of the filament spine, with one side moving away and the other towards. To distinguish between regions based on their average redshift, we label the region with the greater value as region A, and the region with the smaller value as region B. If the galaxies in region A are redshifted ($Z_{A} > Z_{0}$) relative to the host filament, while the galaxies in region B are blueshifted ($Z_{B} < Z_{0}$), then we can infer that the filaments exhibit rotational motion. The redshift difference between the two regions, calculated as:
\begin{equation}
    \rm \dzab = Z_{A} - Z_{B},
\label{equ:dzab}
\end{equation}
 can be used as an indicator of filament spin. The corresponding relative speed is then given by $\mu=c\times\dzab$, where $c$ is the speed of light.

To determine whether the observed $\dzab$ value for a specific filament is statistically significant or could result from random variations, the following step-by-step procedure is performed.
\begin{itemize}
    \item Randomly shuffle the redshifts of the galaxies within a given filament, while keeping their original angular positions fixed. This ensures that the galaxies remain assigned to either region A or region B as originally specified.
    \item Calculate $\dzabr$ of the random trial.
    \item Repeat this random shuffling process 10,000 times for each filament.
    \item Calculate the significance of observed $\dzab$ of each filament using $\rm s=\frac{\dzab-\langle\dzabr\rangle}{\sigma(\dzabr)}$, where $\langle\dzabr\rangle$ and $\sigma(\dzabr)$ are the mean value and standard deviation of those 10,000 random trials.
    \item Calculate the significance of sample-to-sample between selected filaments and their corresponding random trials. 
\end{itemize}

Similarly to the assessment of galaxy rotational characteristics, galaxies can predominantly be categorised into two primary types: dynamically cold or dynamically hot. These correspond to spiral and elliptical galaxies, respectively, and are distinguished by the dynamic temperatures of their star components. We can apply a similar approach to filaments to determine whether a filament is dynamically hot or cold, we examine the redshifts of all galaxies in each filament and compute the root mean square ($\zrms$). We then compare this value with the filament spin signal $\dzab$ for classification. Filaments with $\zrms \textgreater \dzab$ are classified as dynamically hot, while those with $\zrms$ values lower than the redshift difference $\dzab$ are considered dynamically cold. By definition, it is difficult to detect rotation signals in hot filaments, as it is difficult to attribute physical meaning to a $\dzab$ smaller than $\zrms$. Thus, we do not expect to detect any significant rotation signal in dynamically hot filaments.

\section{Result}
\label{sec:result}

\begin{figure*}[!ht]
\plottwo{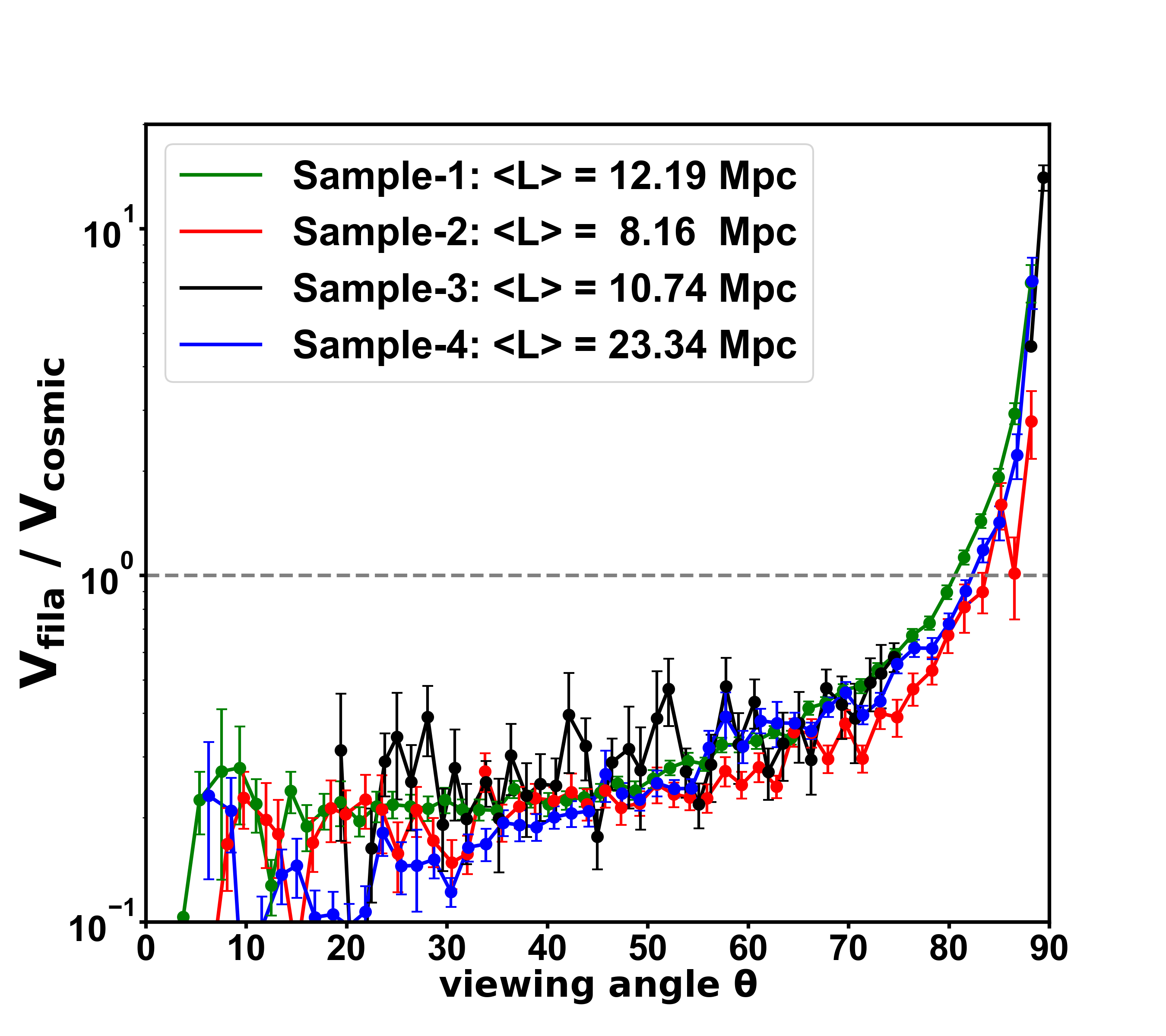}{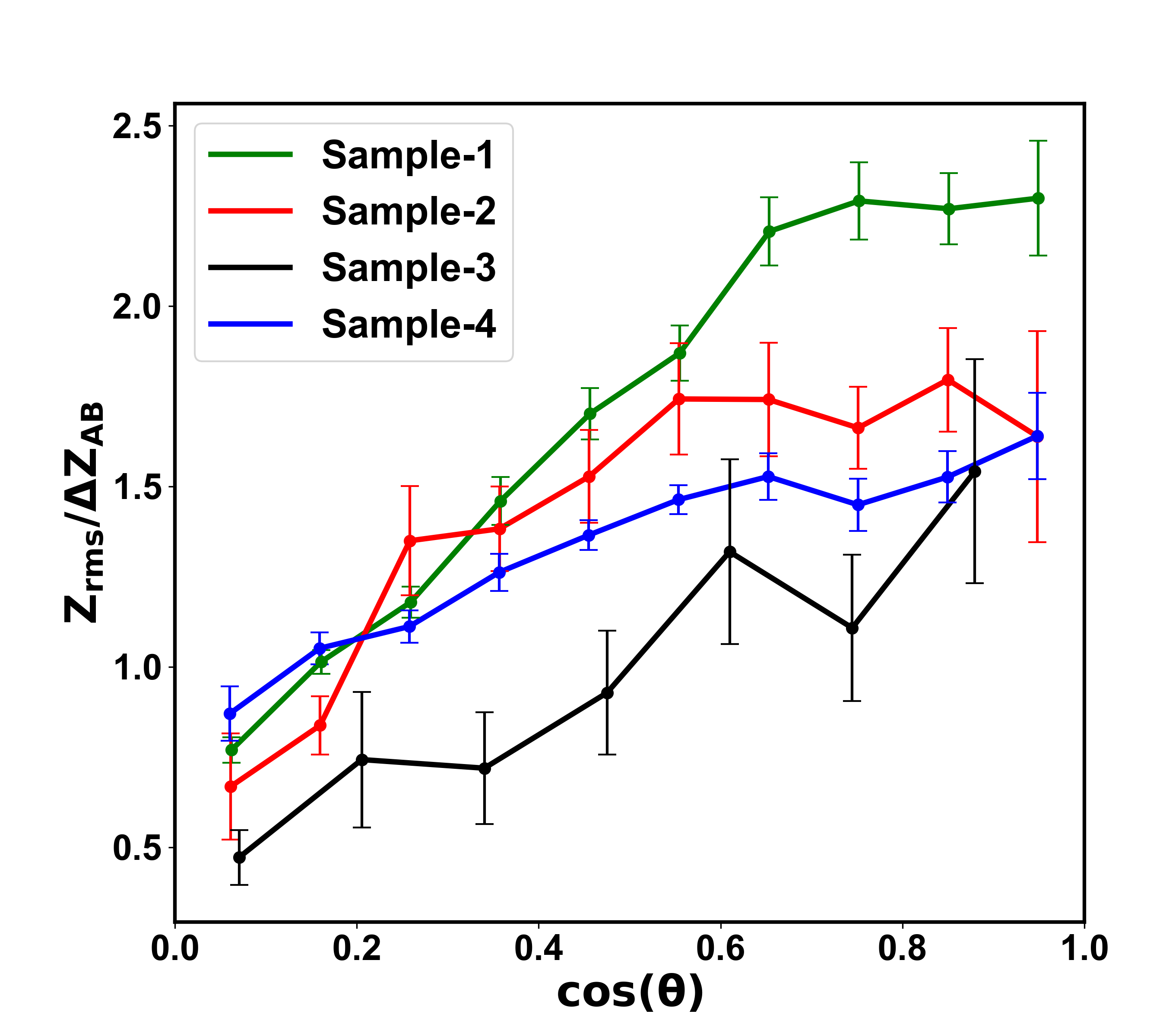}
\caption{\textbf{The correlation between the filaments spin signals ($\dt$ and $\rm V_{fila}$) and the viewing angle ($\rm \theta$) of filament spine.} \textbf{Left panel: the evaluation of filament spin signals in relation to viewing angle.} Here $\rm V_{fila} = c \times \dzab$ is measured signal by our method, $\rm V_{cosmic} \sim H \cdot L \cdot \cos(\theta)$ is the spin signal of combined effects of various observational effects casued by the filament's inclination (see text for more details). A ratio of less than 1 indicates that $\rm V_{fila}$ is unable to effectively distinguish in such cases. The black line (Sample-3) breaks between 75 and 88 degrees is due to missing data because the sample size is too small.
Right panel: The filament dynamical temperature, $\rm Z_{rms}/\dzab$, as a function of the viewing angle, $\theta$, between the filament spine and the line of sight. For a given filament, $\zrms$ is the mean root square of redshifts of all member galaxy and $\dzab$ is the redshift difference of galaxies located on opposite sides of the filament spine. The four colour lines represent different samples, as indicated in the legend. The error bars are determined by 1,000 times of random resampling via bootstrap method.}
\label{fig:dzab_cosine}
\end{figure*}

Filaments are elongated structures made up of numerous galaxies, oriented at different angles relative to the observer's line of sight. Therefore, the first step is to examine how the viewing angle of the filament spine influences the measured filament spin signal.
Consider a scenario where the filament's spine is perfectly aligned with the observer's line of sight, such that galaxies at one end are closer to the observer, while those at the opposite end are farther away.
In this case, the measured signal is proportional to the filament length, $\rm L$, given by 
$\rm V_{cosmic} \sim H \cdot L \cdot \cos(\theta)$.
While base on our method, the filament spin signal is measured approximately by $\rm V_{fila} \sim c \times \dzab$. 
In this configuration, it becomes challenging to determine whether the observed signals are due to the filament's rotational motion or a nonzero signal $\rm V_{cosmic}$ of combined effects of various observational effects caused by the filament's inclination.

In order to address this, we make a rough estimation to examine the ratio between $\rm V_{fila}$ and $\rm V_{cosmic}$ a function of the viewing angle $\rm \theta$ between filament spine to the line-of-sight, which is shown in the left panel of Figure~\ref{fig:dzab_cosine}. Here we adopt the mean value of filament length of each sample. 
It can be seen that the ratio increases with the angle. When the angle is less than approximately 80 degrees, the ratio is less than 1, indicating that the signal measured by our method is weaker than the combined effects of various observational effects. However, when the angle exceeds approximately 80 degrees, the ratio reaches around 1 and rises rapidly. From this, we can conclude that when the angle of the filament spine close to perpendicular to the line-of-sight, the spin signal measured by our method can be distinguished from false signals caused by observational effects.

In the right panel of Figure~\ref{fig:dzab_cosine}, we also show that the dynamical temperature of the filament, $\dt$, is examined as a function of the viewing angle of the filament, $\phi$. The intrinsic dynamic temperature of filaments, if calculated from the motion of galaxies, is independent of the viewing angle. However, because of observational effects, as the figure shows, the dynamical temperature of the filament $\dt$ increases as the filament spine becomes parallel to the line of sight. The same trend for all four samples. The filaments in Sample-2 and Sample-4 (represented by the red and blue lines) show the most comparable behaviour, while Sample-1 (indicated by the green line) shows a higher value of $\dt$ when the filament spine is almost aligned with the line of sight. Interestingly, the dynamic temperatures of the filaments in Sample-3 (shown in black line) are lower than one (which means dynamic cold) at angles less than 45 degrees. 

Two principal factors are postulated to explain this discrepancy. Firstly, variations in the filament search algorithms are considered. Secondly, discrepancies arise because of the differing counts of member galaxies within the filaments. We prefer the latter one, as the increase in the number of member galaxies within a filament correlates with the improved precision in the computation of $\dt$. The higher the number of the member galaxies, the higher the reliability of the derived value $\dt$.

\begin{figure*}[!ht]
\includegraphics[width=0.98\textwidth]{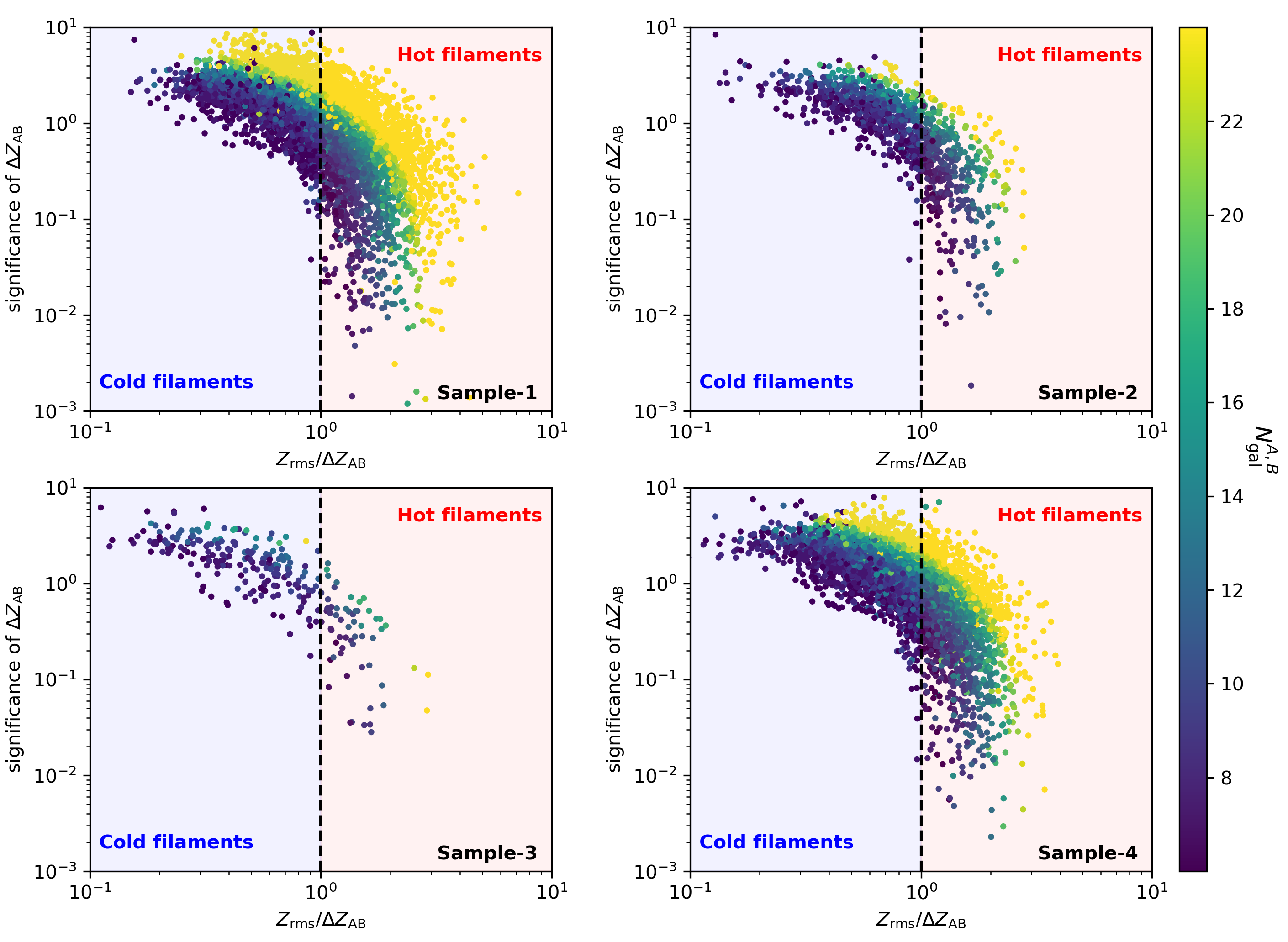}
\caption{\textbf{The comparison of statistical significance of filament spin among four samples, as indicated in the corresponding legends.} The significance of $\dzab$, in units of the standard deviation of 10,000 random trials, for a given filament is shown as a function of the filament's dynamical temperature, $\dt$, where $\dzab$ denotes the redshift difference of galaxies between regions A and B, and $\zrms$ represents the root mean square (r.m.s) of all member galaxies within the filament. A higher value of this quantity suggests that $\dzab$ is less likely to be a random event. The dashed line, $\dt=1$, classifies filaments as either 'cold' or 'hot'.  Each colour-coded point indicates the galaxy richness within a single filament, and yellow denotes filaments with high galaxy richness, while dark purple indicates those with low richness, as shown by the colour bar. $\rm N_{gal}^{A,B}$ is the galaxy number in region A and B.}
\label{fig:sig}
\end{figure*}

As mentioned above, for each filament, galaxies can be divided into two regions (either A or B) based on their spatial location. We then calculate the mean redshift difference ($\dzab$, refer to Equ.~\ref{equ:dzab}) of galaxies between these two regions, which as a proxy of the filament spin. To determine whether this signal is random, we outline the procedure for a random test in Section~\ref{subsec:Spin}. In Fig.~\ref{fig:sig}, we present the comparison of the statistical significance of the filament spin signal among four samples. 
The significance of the spin signal, $\dzab$, of each filament is shown as a function of the dynamic temperature of the filament, $\dt$, in which $\zrms$ is the root mean square (r.m.s.) of all member galaxy of a given filament. The significance of a given filament is calculated in units of the standard deviation of 10,000 random trials (see Section~\ref{subsec:Spin} for more details). The higher value of significance indicates that the spin signal of the filament $\dzab$ is more inconsistent with the random signal.
The filaments are subdivided into dynamic cold or hot according to their dynamic temperature, $\dt$, as shown by the dashed line. The number of galaxies in each region is denoted by colour. Note that the galaxy number ratio between two regions is close to one.

Three notable points can be gleaned from Fig.~\ref{fig:sig}. Firstly, the significance of $\dzab$ decreases with increasing dynamic temperature $\dt$. For the same number of galaxies, the smaller the dynamic temperature of the filaments, i.e., the colder they are, the higher the significance in their spin signal. Secondly, given a fixed dynamic temperature, a higher significance level is associated with a higher count of galaxies within the filament. The third and most critical point is that the general trend remains consistent across all four samples. Different filament algorithms and galaxy samples have the ability to qualitatively reproduce observational findings \citep{2021NatAs...5..839W}, suggesting that a reliable filament spin signal is detectable.

Although the four samples are similar in trend, the differences are noticeable. Samples-1 and sample-4 show a stronger dependence on the number of galaxies. Sample-2 and sample-3, on the other hand, are less distinct. The main reason is the difference in the number of filaments and the number of galaxies in the filaments, which is directly caused by the galaxy samples and filament algorithms.

\begin{figure*}
\includegraphics[width=0.25\textwidth]{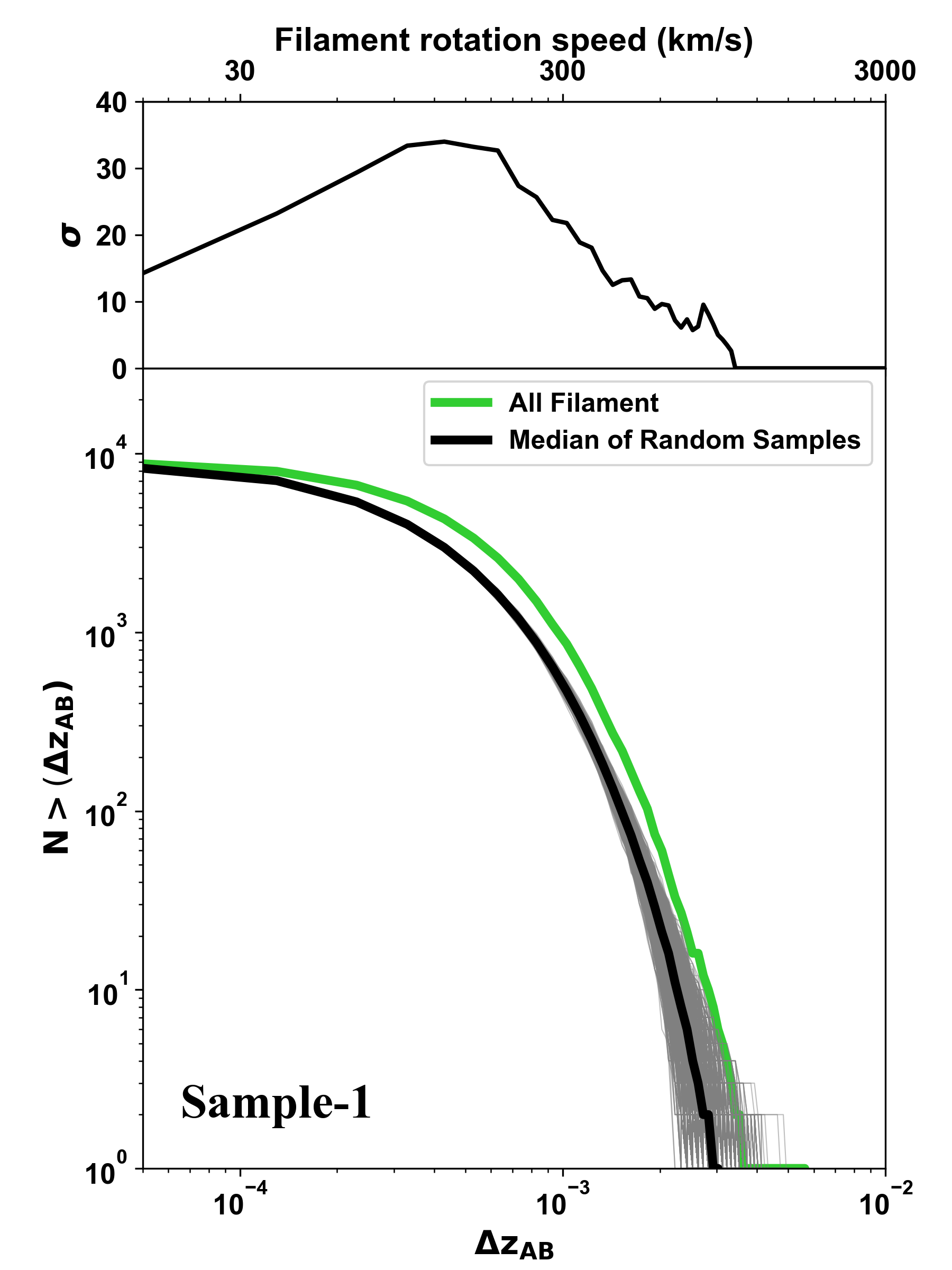}\includegraphics[width=0.25\textwidth]{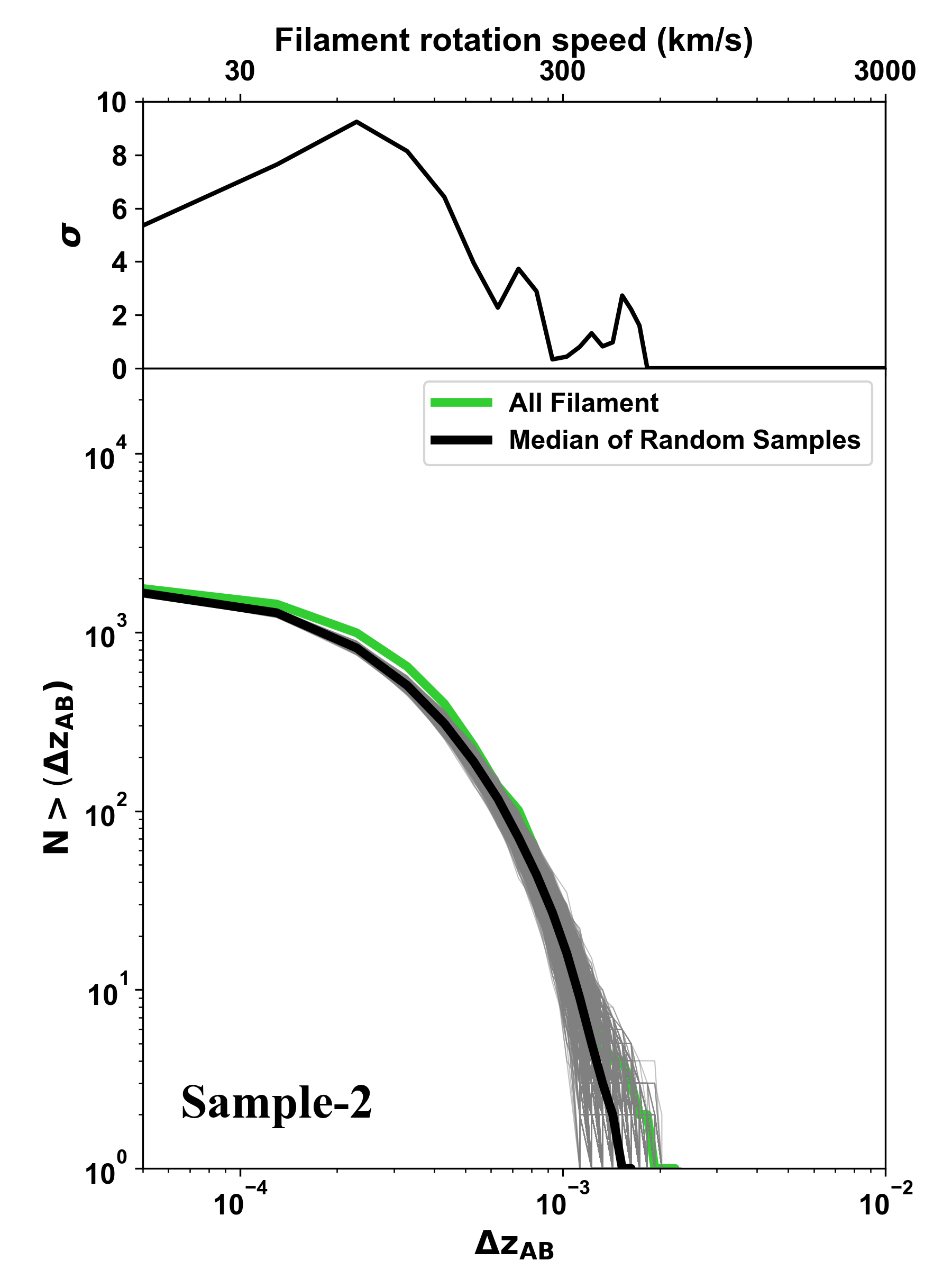}\includegraphics[width=0.25\textwidth]{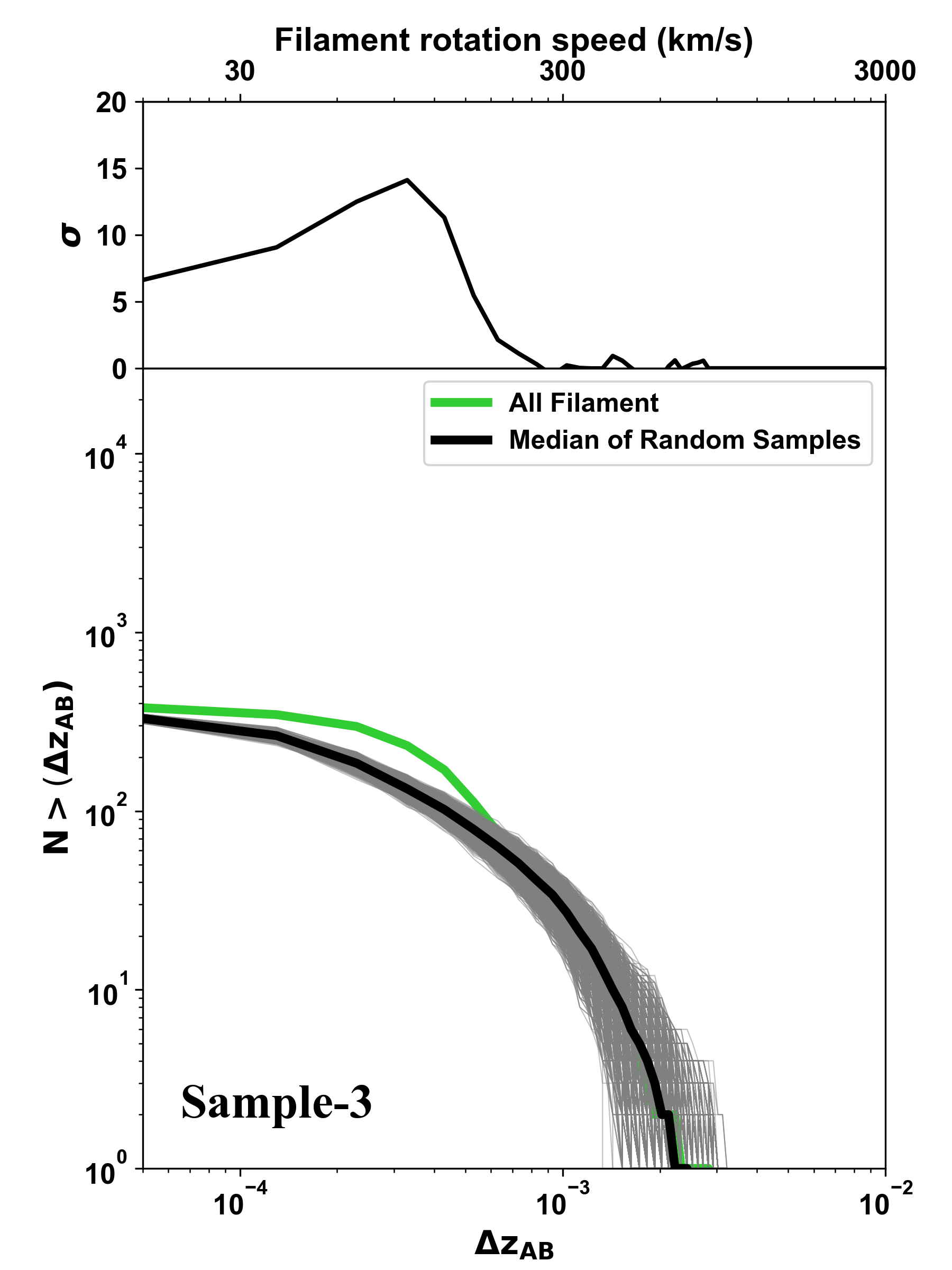}\includegraphics[width=0.25\textwidth]{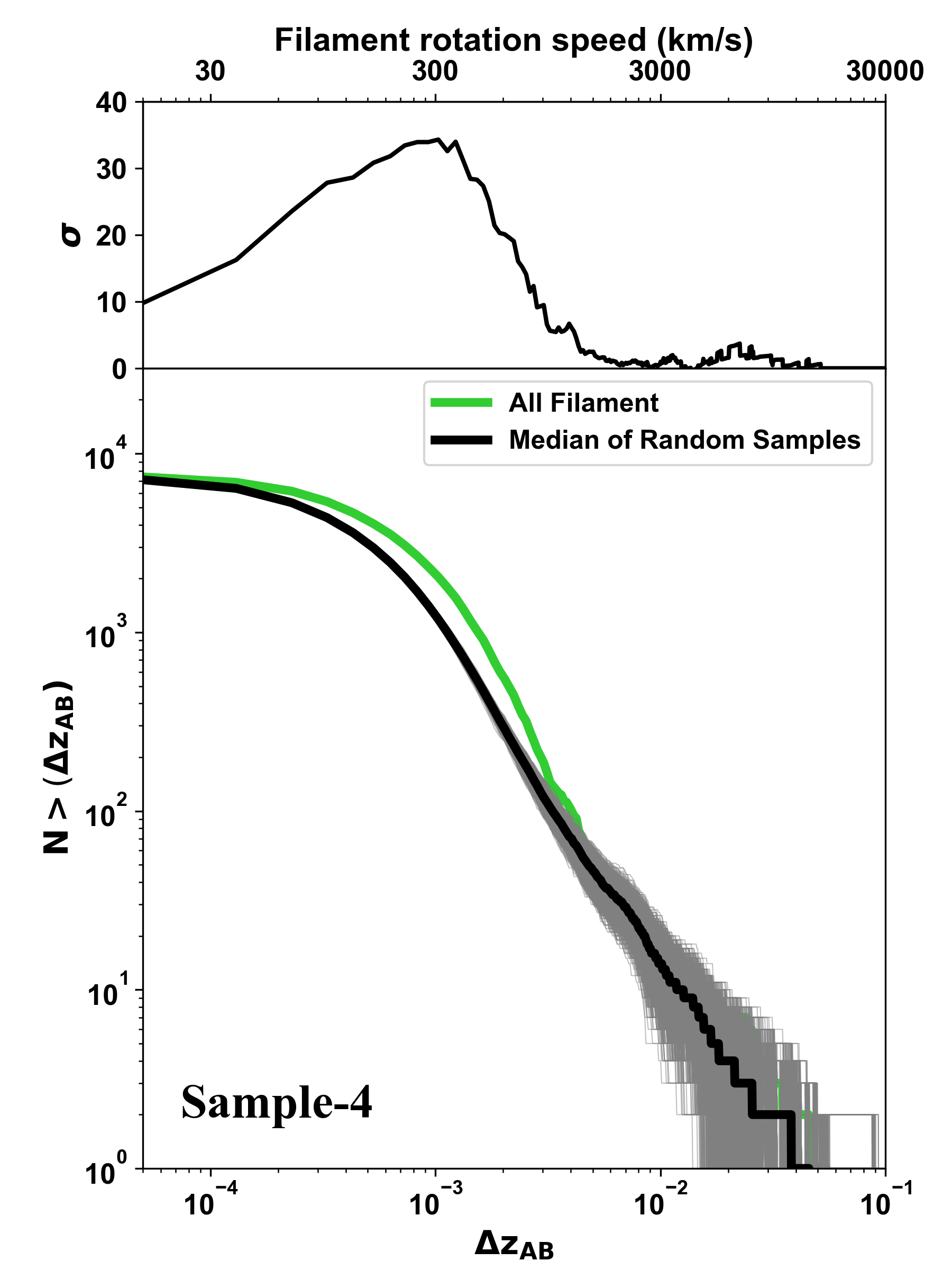}
\includegraphics[width=0.25\textwidth]{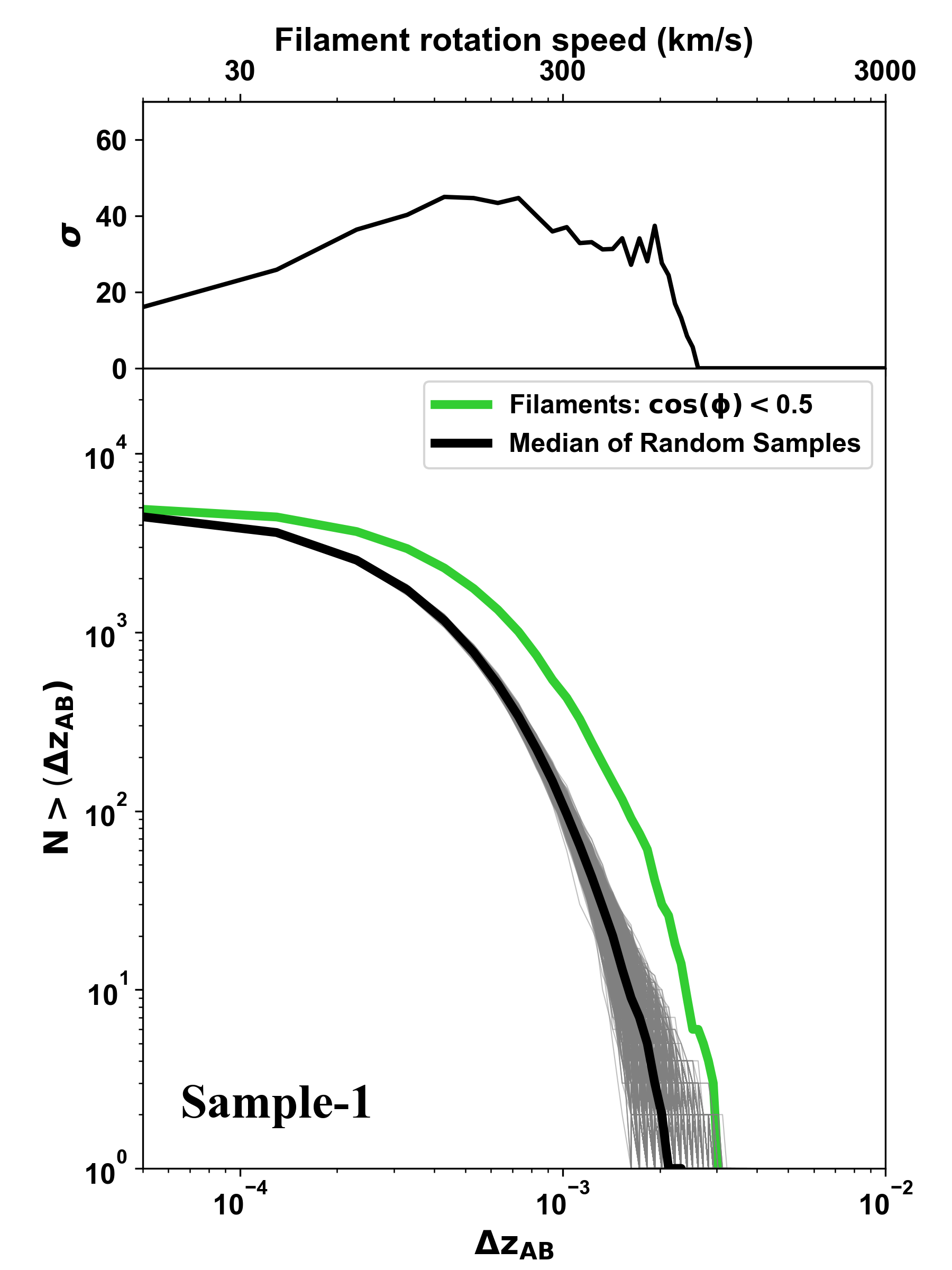}\includegraphics[width=0.25\textwidth]{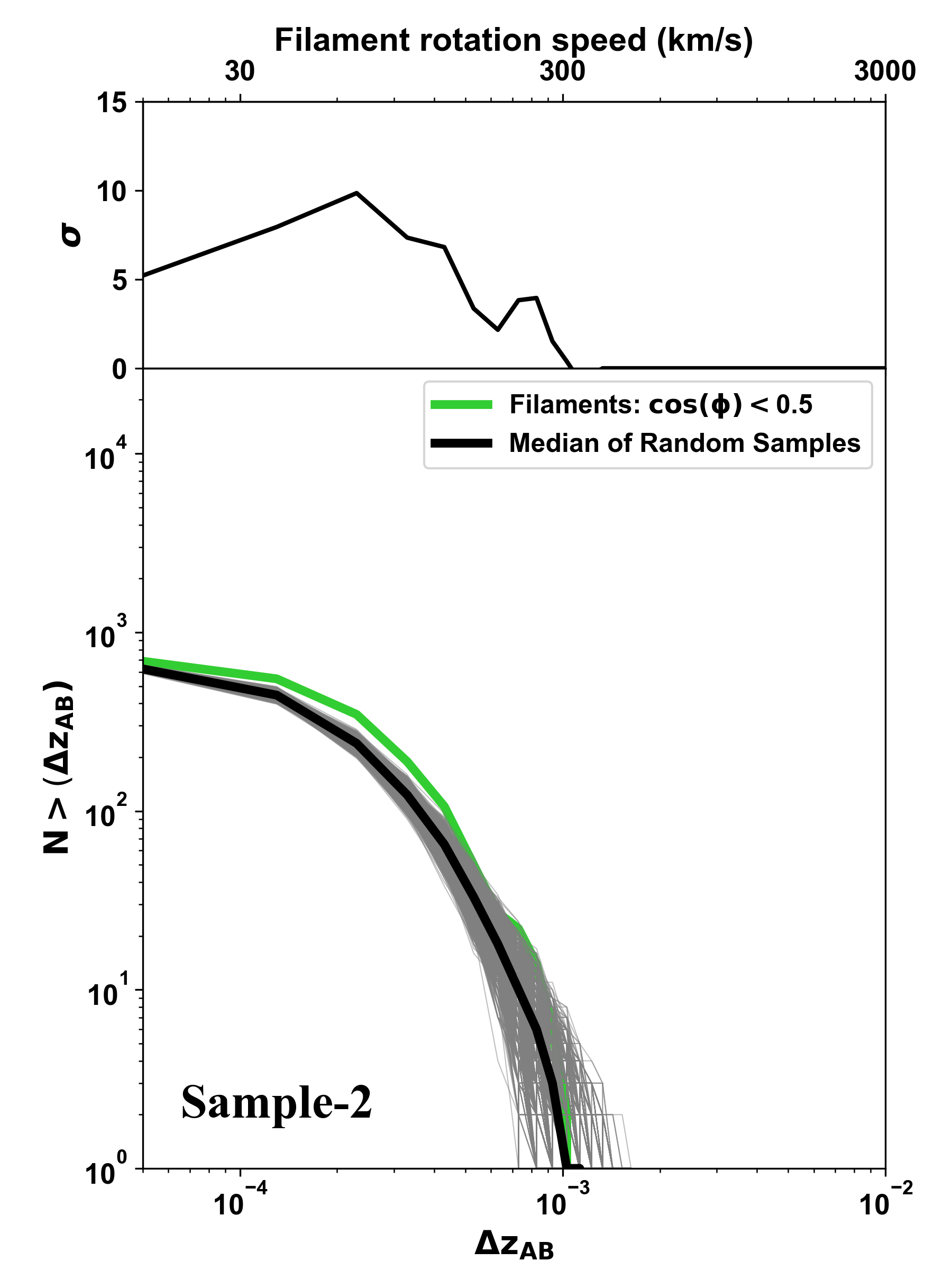}\includegraphics[width=0.25\textwidth]{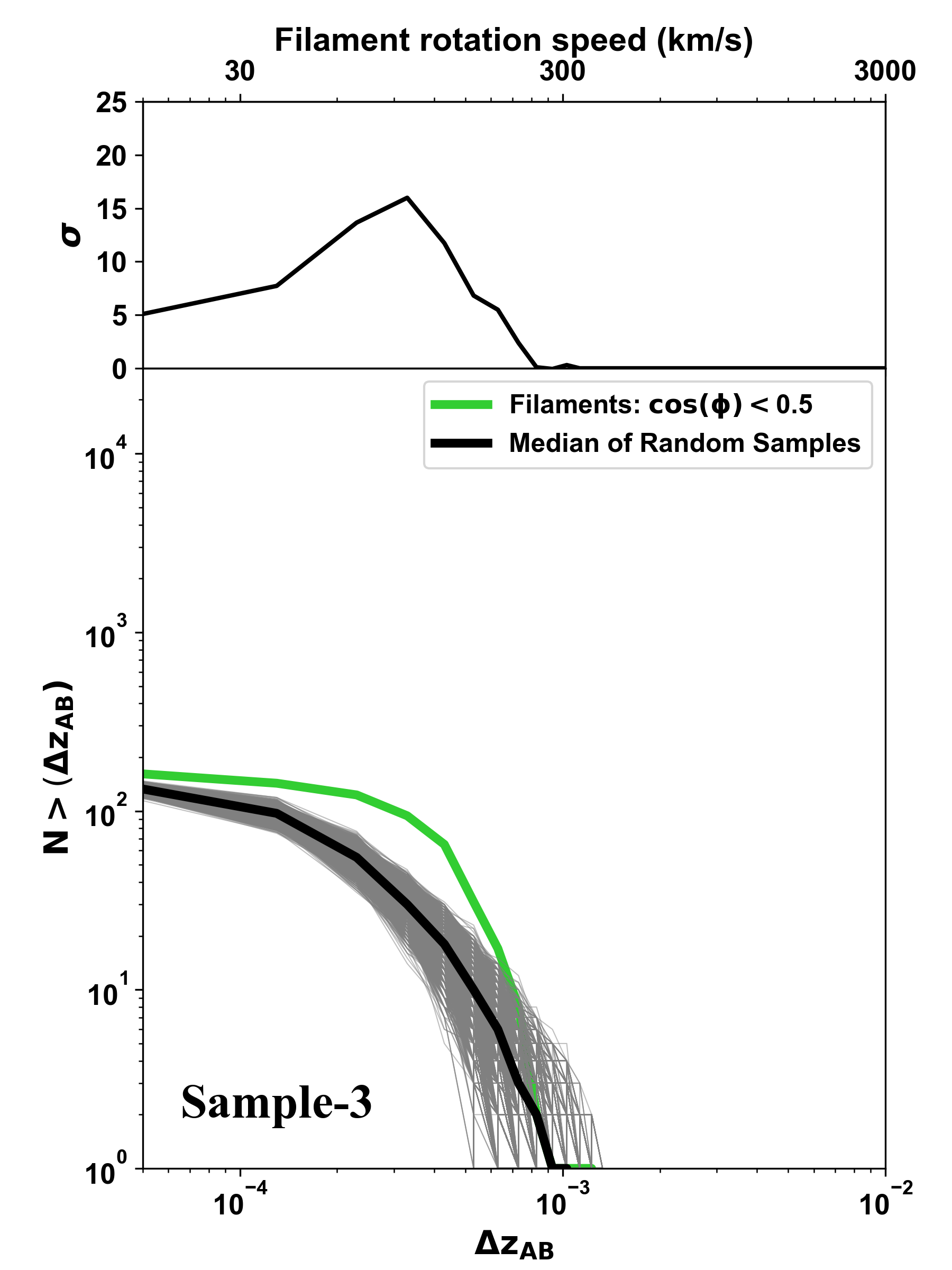}\includegraphics[width=0.25\textwidth]{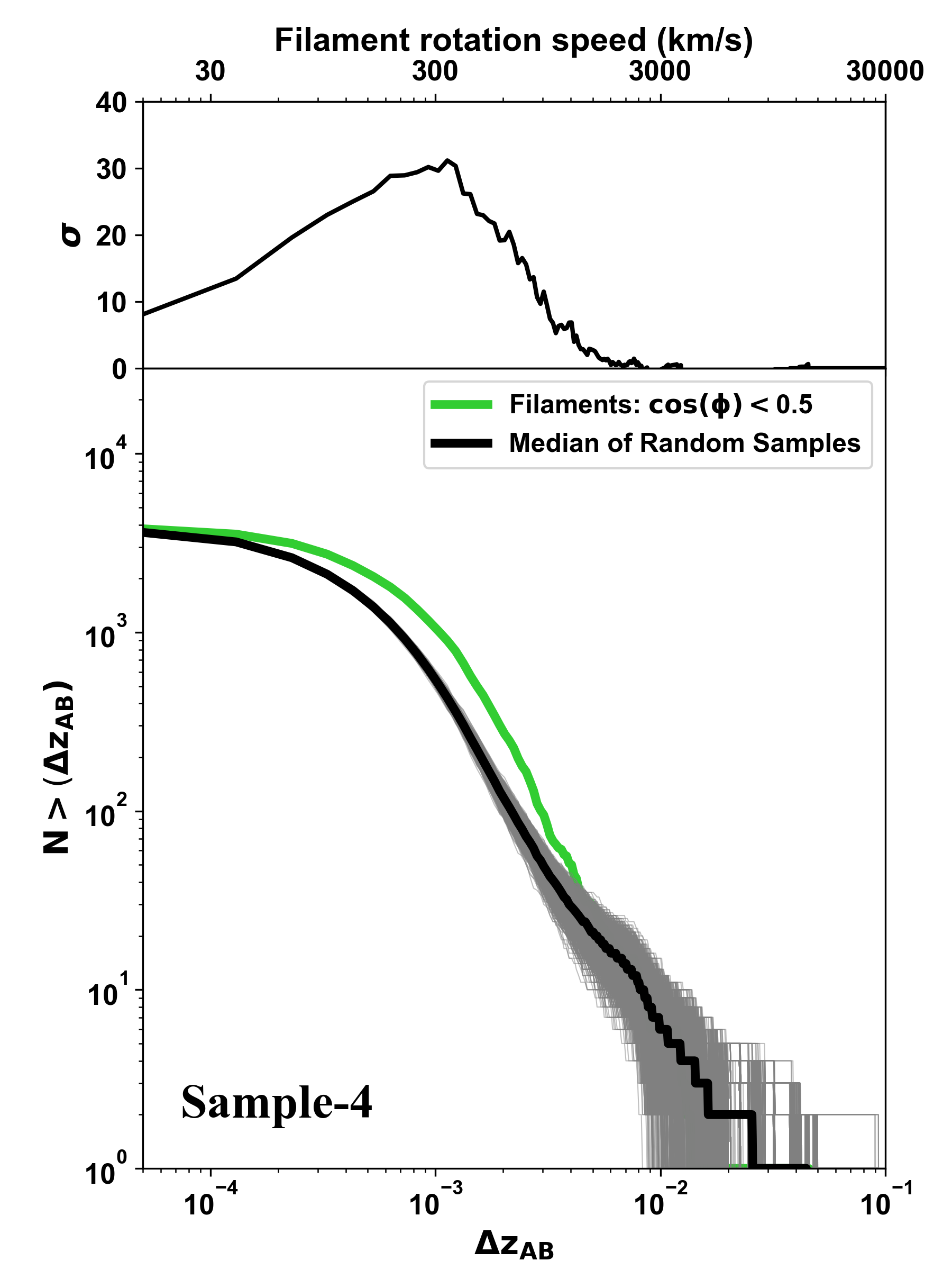}
\includegraphics[width=0.25\textwidth]{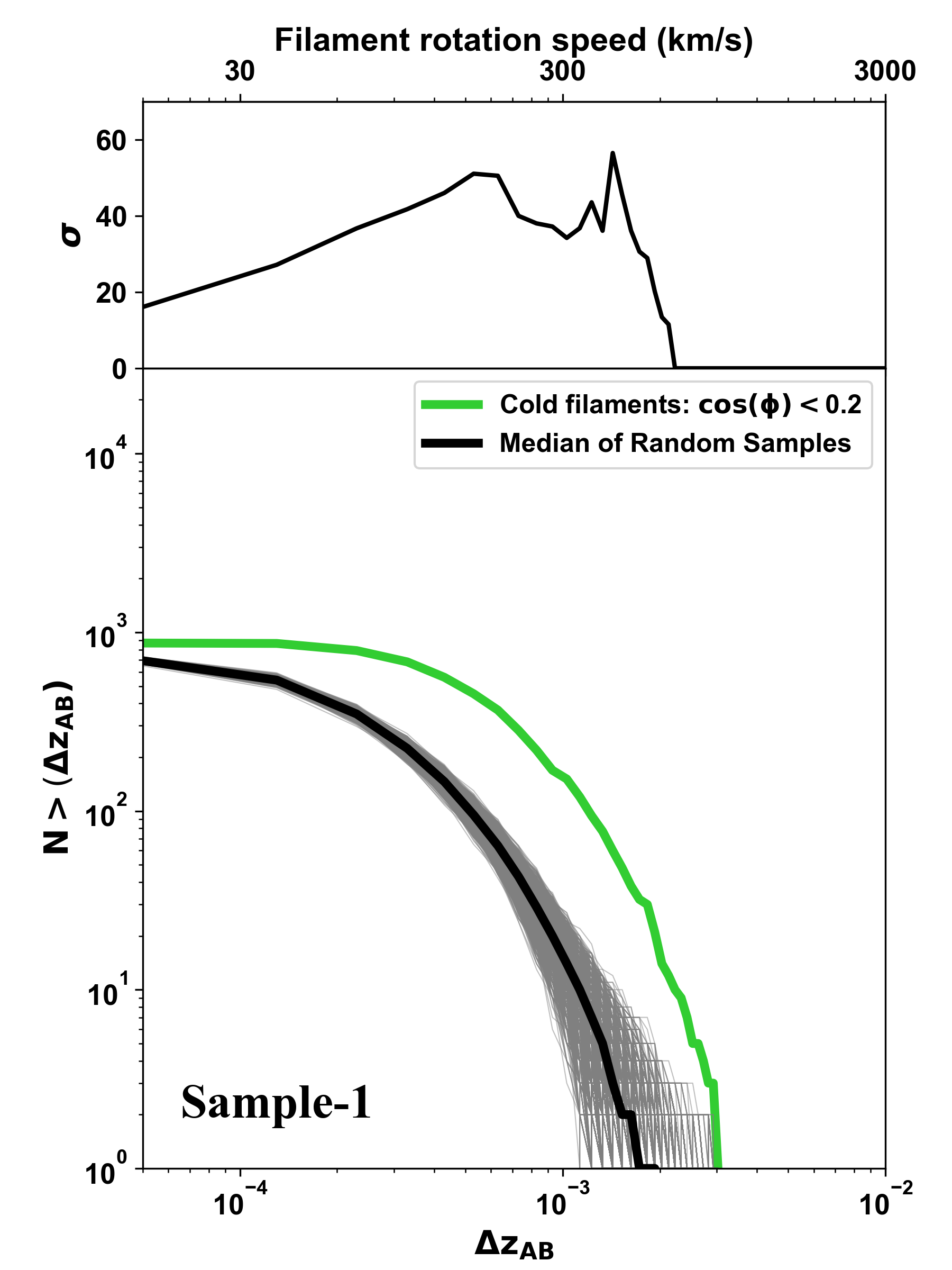}\includegraphics[width=0.25\textwidth]{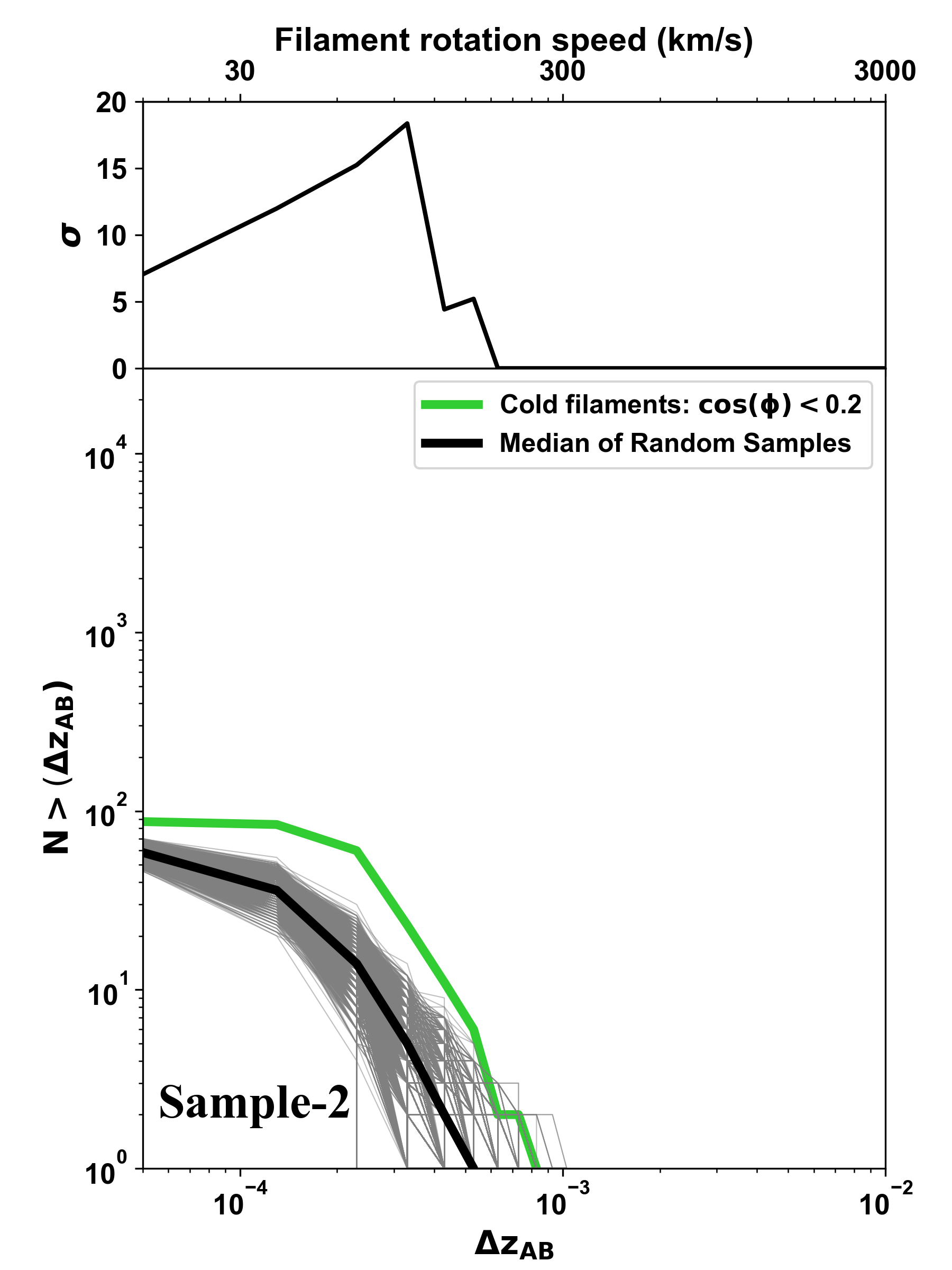}\includegraphics[width=0.25\textwidth]{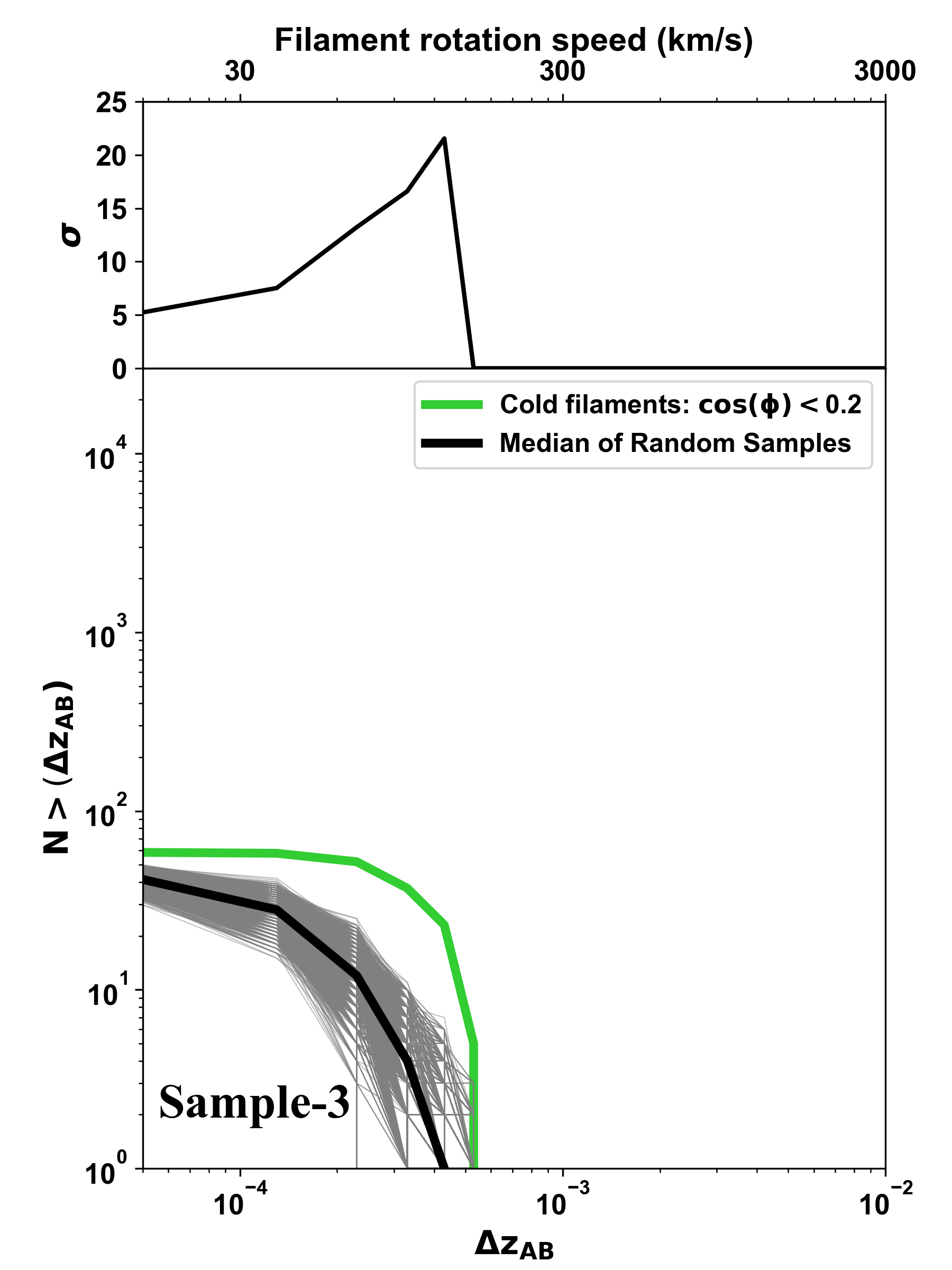}\includegraphics[width=0.25\textwidth]{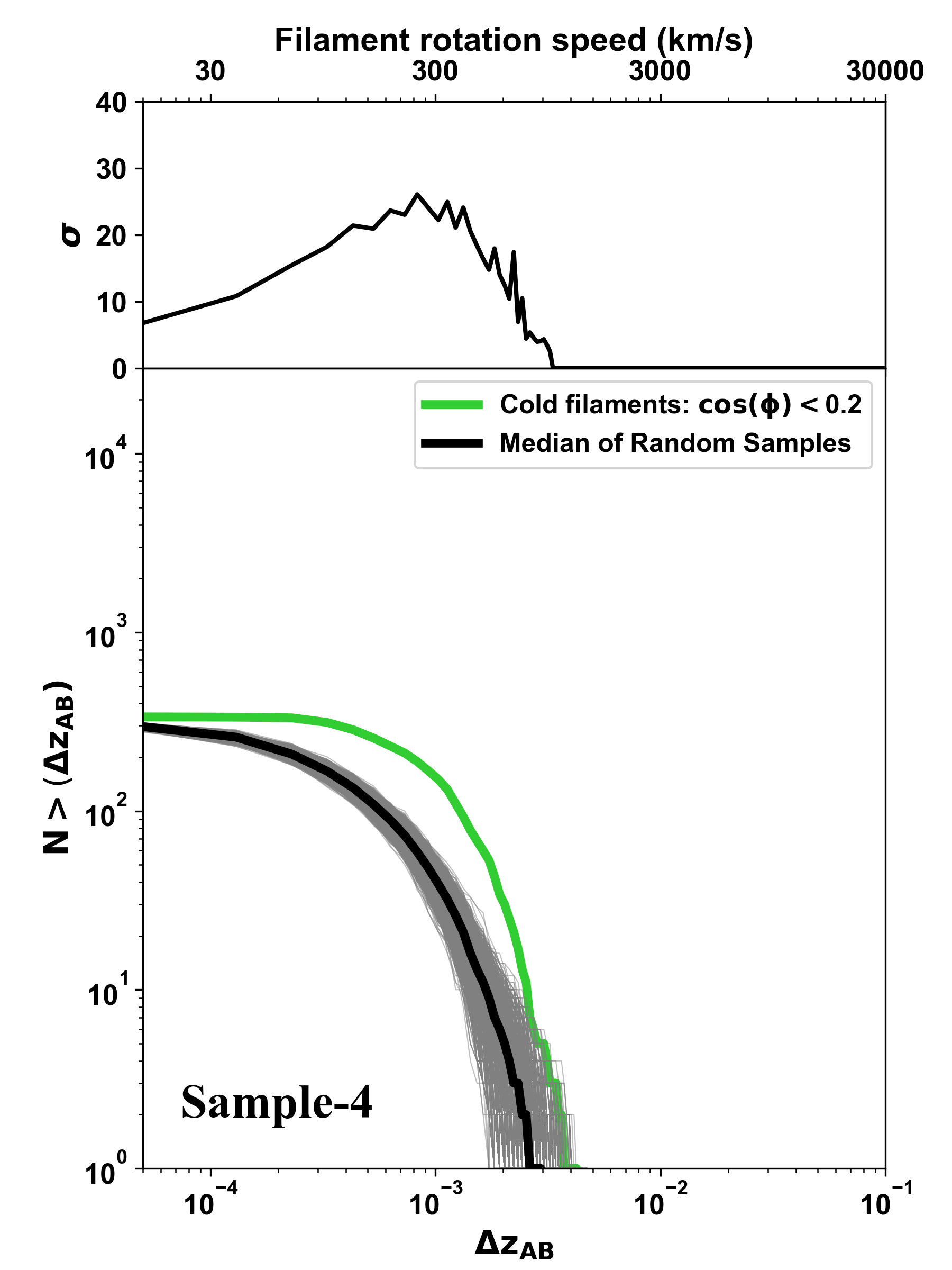}
\caption{\textbf{The spin signals of the filaments compared in four different samples.} Upper panels: considering all filaments. Middle panels: considering filaments with the viewing angle with $\cos(\phi)<0.5$. Bottom panels: considering cold filaments with the viewing angle with $\cos(\phi)<0.2$. Each individual figures from left to right represents one of the samples, as indicated in the legend. The lower sub-figures in each individual figures show the cumulative distribution of redshift differences, $\dzab$, between regions A and B of the filaments. Green lines show the observed filament signals, while the 10,000 grey lines represent the distribution of random samples after randomly shuffling the galaxies' redshifts. The solid black line marks the median of these 10,000 randomised samples. The upper sub-figures in each individual figures show the discrepancy, in standard deviation units of the randomised distribution, between the observed curve and the mean of the random distributions as a function of $\Delta z_A$. The top x-axis shows the filament rotation velocity in $\rm km/s$, calculated as $\mu=c\times\dzab$. }
\label{fig:dzab_all}
\end{figure*}

Subsequently, our attention is move to focus on examining the spin signals of the filaments, denoted as $\dzab$, which provides insightful information on their rotational characteristics. 
In Figure~\ref{fig:dzab_all}, we present the spin signals of the filaments. Each of these figures contains four sub-figures, with each sub-figures corresponding to one of four samples (which are clearly marked in the legend located in the left corner). Note that top panels ignores the viewing angle between the filament spine and the line-of-sight. The influence of the viewing angle on the signals is shown in the middle and bottom panels. The bottom sub-figures in each panel displays the cumulative distribution function (CDF) of the observed spin signals $\dzab$ using a solid green line, while random samples are represented by multiple solid grey lines. In addition, the mean value of these random samples is indicated by a solid black line. The top panels present the deviations of the measured samples from their corresponding random trials, allowing a clear comparison of the discrepancies between the measured signal and the randomised samples.

All four sample sets show a qualitatively significant and consistent signal aligned with the observational results reported by \citet{2021NatAs...5..839W}. 
Among the four sets of samples, sample-2 exhibited the faintest signal, whereas sample-1 displayed a comparatively strong signal. 
The overall trends of filament spin signals regarding the viewing angle for a given sample align with the findings of \citet{2021NatAs...5..839W}, for which filament spine perpendicular to the line-of-sight show relative strongest signal. 
From top to bottom, in each sample, as the considered viewing angle approaches a more perpendicular configuration, the measured signal (green line) deviates further from the gray line, which represents the random sample. This indicates, as discussed above, that larger viewing angles make the filament spin signal easier to detect, which agree with what we show in the left panel of Figure~\ref{fig:dzab_cosine}.
This consistency remains intact even though, in certain cases, specific samples tend to obscure the random signal at higher $\dzab$ values. Moreover, within the range of approximately $10^{-4}$ to $10^{-3}$, each set of samples exhibits a signal significance that almost reaches $10\sigma$, underscoring the robustness of the signal observed of filament spins.

\begin{figure*}
\includegraphics[width=0.98\textwidth]{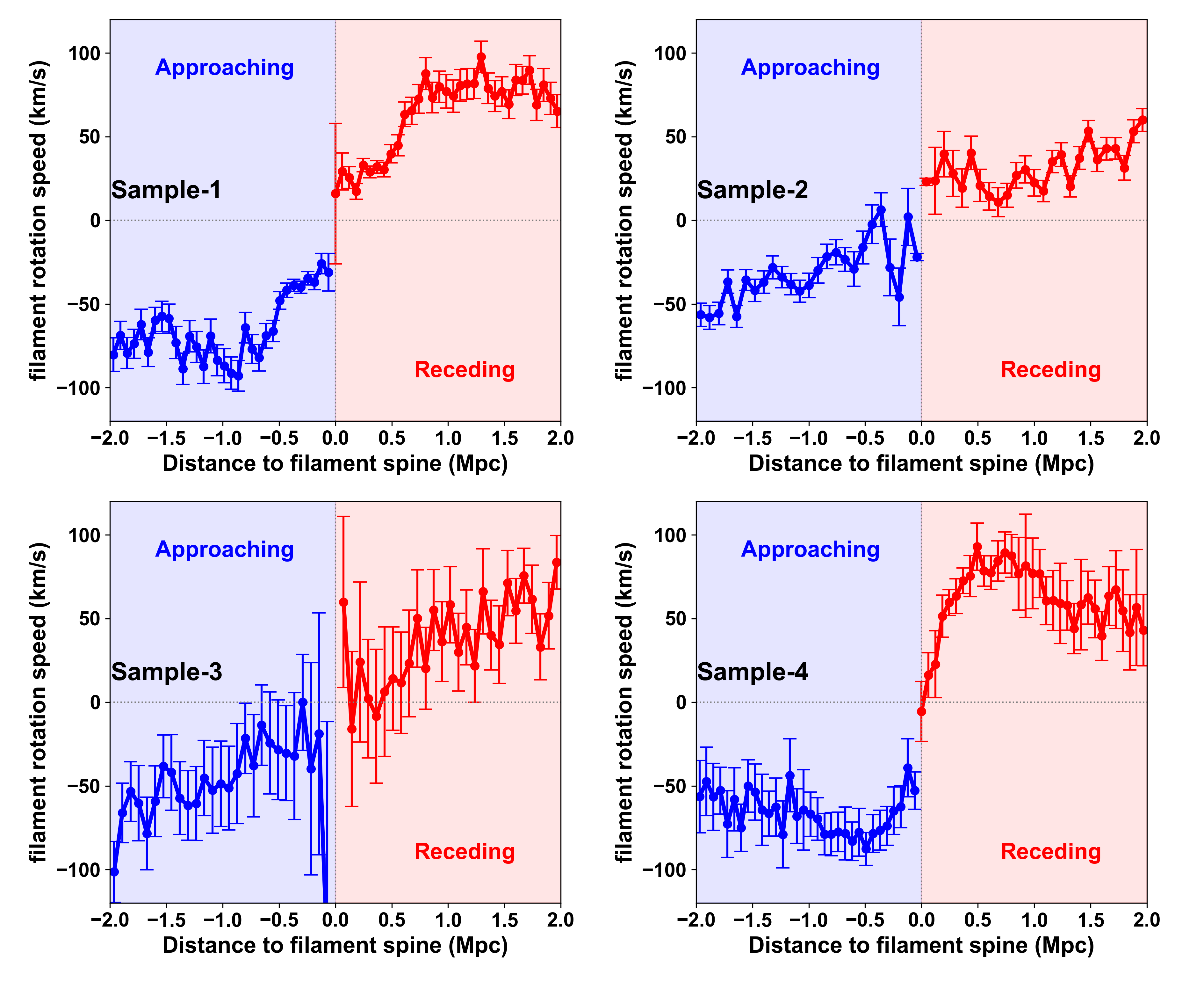}
\caption{\textbf{The rotation curve of filaments in four samples, as indicated in the legend.} The rotational speed of the filament is examined based on the distance between galaxies and the filament spine. The rotation speed is computed using $c \times \Delta Z$, where $\Delta Z$ represents the difference in the redshift between galaxies at certain distances relative to the redshift of the filament, which is the mean value of all galaxies within the filament. Galaxies that are in regions moving away from the observer are depicted in red, and the distance of these galaxies from the filament spine is assigned positive values. Conversely, galaxies in regions moving towards the observer are shown in blue, with their distances assigned negative values. The error bars represent the standard deviation of the mean.}
\label{fig:rc}
\end{figure*}

In Figure~\ref{fig:rc}, we performed a quantitative analysis of the filament spin signals investigating the rotation curve of the filament spins. This rotation curve was derived for the stacked sample of filaments by converting the observed redshift differences into velocity differences, which are calculated as $\rm c\times\Delta Z$. Here, $\rm \Delta Z$ represents the redshift difference between galaxies at a given distance and the average redshift of all galaxies within the filament, with $c$ denoting the speed of light.  We employed the conventional representation where velocities are classified as negative for motion towards us (approaching, in blue) and positive for motion away (receding, in red). 

In general, the analysis presented in Figure~\ref{fig:rc} illustrates that the rotational speed exhibits an increase as the distance increases, but the differences are also obvious. 
For Sample-1, the rotational velocity of the filaments initially increases with increasing distance, reaching its peak near 1 $\rm Mpc$, after which the speed slightly decreases for distances beyond 1 $\rm Mpc$. 
However, in the filaments of sample-2 and sample-3, the velocities continue to increase as the distance increases, with no indication of a decreasing tendency. For sample-2, the maximum velocity is approximately 50 $\rm km/s$, while for sample-3, it almost reaches 100 $\rm km/s$.
As we mentioned, the overlap of galaxies among samples close to 90\%, the main difference of galaxy sample in sample-2 and sample-3 is that the galaxy redshifts are not corrected for RSD. As suggested by \cite{2024MNRAS.532.4604W} which the RSD will make the filament radius twice from 1 $\rm Mpc$ to 2 $\rm Mpc$, which may lead to a different rotation curve
What shows in the right-bottom panel (Sample-4) demonstrates that its filament's rotational velocity aligns well with \citet{2021NatAs...5..839W}. Initially, the speed increases, peaking around $\rm 100 km/s$ at 1 Mpc, before declining to near $\rm 50 km/s$ at 2 Mpc. However, the observational work \citet{2021NatAs...5..839W} shows a total deceleration to zero at 2 Mpc, marking the main difference with sample-4. 
This discrepancy is attributed solely to the filament algorithm, as it is the only distinction from \citet{2021NatAs...5..839W}.

\section{Summary and Discussion}\label{sec:sum_dis}

Observations \citep{2021NatAs...5..839W} and simulations \citep{2021MNRAS.506.1059X,2022PhRvD.105f3540S} have confirmed the rotational motion of cosmic filaments. However, a comprehensive assessment of whether diverse data sets and algorithms, especially those related to filament identification, can consistently reproduce previous findings is warranted. 
This is the primary motivation for this study, which seeks to understand the potential impacts of data sets and algorithms on the quantification of filament spin.

In this study, we implemented four distinct filament algorithm sets. Additionally, the galaxies used for measuring the spin signal of these filaments varied somewhat in composition. The results obtained are as follows:
\begin{itemize}
    \item Based on a rough estimation, when the viewing angle of the filament spine exceeds approximately 80 degrees, the spin signal measured by our method can be distinguished from nonzeros signals caused by other cosmological effects.
    \item For all filaments in those four samples, the filament spin signal can be reproduced qualitatively with similar results. The spin signal depends on the dynamical temperature $\dt$ and the viewing angle $\phi$ of the filaments, which agrees with the observational results.
    \item Our quantitative analysis revealed a variation in the filament rotation curve. The distance within a range of less than 1 Mpc, previous observational results are broadly replicated as rotational velocity generally increases with distance. However, for a distance larger than 1 Mpc, the situations of the rotation curve varies among samples, with some indicating a continued increase while others exhibit a reduction.
    \item The RSD effect, although it cannot qualitatively change the spin signal of the filaments, can quantitatively affect it to a large extent, especially in the rotation curve where it is most pronounced. 
\end{itemize}

Our results are in qualitative agreement with previous studies \citep{2021NatAs...5..839W, 2021MNRAS.506.1059X, 2022PhRvD.105f3540S}, particularly with the observational analysis presented in \citep{2021NatAs...5..839W}, supporting the idea of filament rotation. This agreement holds across different filament identification algorithms and galaxy samples. However, the observed quantitative differences are mainly attributed to variations in the algorithms and the RSD effect of the galaxy samples. To explore this further, we focused on filaments in the overlap region and used the same galaxy sample to measure the spin signal. Although this approach reduced, but did not fully eliminate, the discrepancies in the distribution of $\dzab$ and the rotation curves, the remaining variations indicate that the differences arise from the filament identification algorithms.

The Zel'Dovich approximation \citep{1970A&A.....5...84Z} provides a theoretical framework for understanding the formation and evolution of cosmic filaments. It also illuminates the dynamics and motion of matter, specifically galaxies, that are associated with these structures. 

Initially, matter undergoes a collapse primarily along the principal compression axis, resulting in the formation of extensive cosmic structures known as walls. Subsequently, matter flows within the plane of these walls, following the direction determined by the intermediate compression axis, thus giving rise to filamentary structures. During this final stage, the matter collapses and moves along the axes of the filaments, leading to the assembly of the matter into clusters \citep{Icke1991,2014MNRAS.441.2923C}. It is important to note that the collapse along the three axes occurs sequentially, proceeding at varying velocities.

In this model of mass flow, the spin of the filament is expected to emerge during the collapse process, driven by tidal torques and the anisotropic collapse of matter around the filament. In terms of galaxy dynamics, while some galaxies can move directly into clusters from voids and do not contribute to filament formation, those galaxies that are hosted by filaments are accreted in a two-phase process \citep{2018MNRAS.473.1562W}. First, they move toward the filament from the surrounding wall, and then they flow along the filament toward the cluster. This pathway is consistent with the complex structure formation in the cosmic web. Assuming filaments can be approximated by a cylindrical potential, the motion of galaxies—particularly in a helical pattern—can naturally give rise to a rotation around the filament spine.

However, the real situation is more complex. Filaments are not perfectly cylindrical in distribution, and the algorithms for the identification of filaments are unable to identify a perfect cylindrical structure.
The helical motion of galaxies varies at different places in the filament (i.e., some are already dominated by motion along the filament to the cluster, while others are still dominated by vertical infall towards the inner region of the filament).
Filaments are frequently perceived as composites of several segments, with these segments exhibiting minimal deformation, indicating that the filament spine remains relatively stable. This structural characteristic primarily accounts for the observed relationship between the detected rotational signal, the dynamic temperature of the filament, and the viewing angle.

The Tidal Torque Theory (TTT) has been relatively successful in explaining the generation of angular momentum on scales such as dark matter halos and galaxies. However, its application to cosmic-scale structures like filaments, particularly during the transition from quasi-linear to nonlinear regimes, remains less explored. Numerical simulations have shown that collisionless cold dark matter particles can generate spin in cosmic filaments \citep{2020OJAp....3E...3N,2021MNRAS.506.1059X,2022PhRvD.105f3540S}, but the underlying physical mechanisms driving this process at these larger scales warrant further investigation.

Here, as the first step of the \textit{Cosmic Filament Spin} project, we have examined filament signals in redshift space. We acknowledge that asymmetrical redshift-space filament shapes not completely arising from spin (present, e.g. even in real space) might contribute in some cases to the observed spin signal. A detailed investigation of this influence, including an evaluation of possible false positives, will be addressed in a forthcoming paper \citep{2025ApJ...983..100W}. Also, much further study is necessary of the origin and evolution of filament spin, and how consistent the observations are with existing theoretical ideas.

\bigskip

\textbf{Acknowledgments}

The authors appreciate Agust{\'\i}n Rost for sharing the filament \& galaxy catalogue which was used in his work \citep{2020MNRAS.493.1936R}.
PW sponsored by Shanghai Pujiang Program(No.22PJ1415100). 
XXT and PW acknowledge financial support by the NSFC (No. 12473009). XXT and PW also sponsored by Shanghai Rising-Star Program (No.24QA2711100).
HX acknowledges financial support by the NSFC (No.12403010).
%

\bibliography{main}{}
\bibliographystyle{aasjournal}




\end{CJK*}
\end{document}